\documentclass[12pt, a4paper]{article} 
\usepackage{amsmath,amssymb,bm,epsfig,afterpage}
\usepackage{cite}
\usepackage{here}
\usepackage{color}
\usepackage{colortbl}
\usepackage{tabularx}
\usepackage{subcaption}
\usepackage{bbm}
\usepackage{graphicx}
\usepackage{listings}
\usepackage{longtable}
\usepackage[utf8]{inputenc}

\definecolor{Orange}{cmyk}{0,0.61,0.87,0}
\definecolor{JungleGreen}{cmyk}{0.99,0,0.52,0}
\definecolor{OliveGreen}{cmyk}{0.64,0,0.95,0.40}
\definecolor{Brown}{cmyk}{0,0.81,1,0.60}
\definecolor{RoyalBlue}{cmyk}{0.71,0.53,0,0.12}
\definecolor{Gray}{cmyk}{0,0,0,0.40}
\definecolor{LightPink}{cmyk}{0.0,0.25,0,0}
\definecolor{LLightPink}{cmyk}{0.0,0.10,0,0}
\definecolor{LightBlue}{cmyk}{0.25,0,0,0}
\definecolor{LightGray}{cmyk}{0,0,0,0.2}

\newcommand{\vev}[1]{{\left\langle{#1}\right\rangle}}
\newcommand{\br}[2]{\text{Br}({#1}\to{#2})}  
\newcommand{\abs}[1]{\left|{#1}\right|}
\newcommand{\order}[1]{\mathcal{O}\left( {#1}\right)}
\newcommand{\la}{\lambda}

\newcommand{\eps}{\epsilon}
\newcommand{\hla}{\hat{\lambda}}

\newcommand{\mueff}{\mu_\mathrm{eff}} 
\newcommand{\Beff}{B_\mathrm{eff}} 
\newcommand{\laeff}{\la_\mathrm{eff}} 
\newcommand{\scal}{\varphi}
\newcommand{\ol}[1]{\overline{#1}} 
 
\newcommand{\Acal}{\mathcal{A}}
\newcommand{\Lcal}{\mathcal{L}}
\newcommand{\Mcal}{\mathcal{M}}
\newcommand{\Ncal}{\mathcal{N}}
\newcommand{\Ocal}{\mathcal{O}}
\newcommand{\tMcal}{\tilde{\mathcal{M}}} 
\newcommand{\Zn}[1]{\mathbbm{Z}_{#1}} 
\newcommand{\Kahler}{{K$\mathrm{\ddot{a}}$hler}}

\allowdisplaybreaks[3]
\newcolumntype{Y}{&gt;{\centering\arraybackslash}X} 
\usepackage[colorlinks=true, linkcolor=OliveGreen, citecolor=RoyalBlue,
urlcolor=RoyalBlue]{hyperref}

\begin{document}
\begin{titlepage}
\begin{flushright}
{\tt 
}
\end{flushright}

\vskip 1.35cm
\begin{center}

{\Large
{\bf
A low-scale flavon model with a $\mathbbm{Z}_N$ symmetry
}
}

\vskip 1.5cm

Tetsutaro~Higaki $^a$\footnote{
E-mail address: 
\href{mailto:thigaki@rk.phys.keio.ac.jp}{\tt thigaki@rk.phys.keio.ac.jp}}
and
Junichiro~Kawamura$^{b,a}$\footnote{
E-mail address: 
\href{mailto:kawamura.14@osu.edu}{\tt kawamura.14@osu.edu}} 

\vskip 0.8cm

{\it $^a$Department of Physics, Keio University, Yokohama 223-8522, Japan} 
\\[3pt]
{\it $^b$Department of Physics, Ohio State University, Columbus, Ohio
 43210, USA} 
\\[3pt]

\date{\today}

\vskip 1.5cm

\begin{abstract}
We propose a model that explains the fermion mass hierarchy 
by the Froggatt-Nielsen mechanism with a discrete $\mathbbm{Z}_N^F$ symmetry.
As a concrete model, we study a supersymmetric model with a single flavon
coupled to the minimal supersymmetric Standard Model. 
Flavon develops a TeV scale vacuum expectation value for realizing flavor hierarchy,  
an appropriate $\mu$-term and the electroweak scale, hence the model has a low cutoff scale.
We demonstrate how the flavon is successfully stabilized 
together with the Higgs bosons in the model.
The discrete flavor symmetry $\mathbbm{Z}_N^F$ 
controls not only the Standard Model fermion masses, 
but also the Higgs potential and a mass of the Higgsino which is a good candidate 
for dark matter. 
The hierarchy in the Higgs-flavon sector is determined 
in order to make the model anomaly-free and realize a stable electroweak vacuum. 
We show that this model can explain 
the fermion mass hierarchy, realistic Higgs-flavon potential 
and thermally produced dark matter at the same time. 
We discuss flavor violating processes induced by the light flavon 
which would be detected in future experiments. 
\end{abstract}

\end{center}

\end{titlepage}

\setcounter{footnote}{0}

\tableofcontents
\clearpage 
\section{Introduction} 
The origin of hierarchical structure of the fermion masses and the CKM matrix 
is a long standing mystery in the Standard Model (SM). 
The Froggatt-Nielsen (FN) mechanism~\cite{Froggatt:1978nt} is known to be one of solutions for this problem.
Based on the mechanism, a singlet field, the so-called flavon, and a flavor dependent extra symmetry are introduced to the SM, 
so that the hierarchy between the Yukawa couplings are explained
by powers of a ratio of a vacuum expectation value (VEV) of flavon to a cutoff scale of a model:
A source for the fermion mass hierarchy is given by
\begin{align}
\eps := \frac{\vev{S}}{\Lambda} \ll \order{1},
\end{align}
where $S$ is a flavon and $\Lambda$ is a cutoff scale.
It is well known that the FN mechanism successfully explains the fermion mass hierarchy 
and the CKM matrix~\cite{Babu:2009fd}.

It is required for realizing the hierarchy that a flavon is a singlet under the SM gauge symmetry but carries a flavor symmetry charge 
and develops a non-zero VEV.
Any particle can be identified as a flavon as far as these properties are satisfied~\footnote{ 
Recently, it is proposed that a flavon can be identified as the QCD axion~\cite{Ema:2016ops,Calibbi:2016hwq,Arias-Aragon:2017eww,Alanne:2018fns,Ema:2018abj,Bonnefoy:2019lsn}. 
See also for earlier works~\cite{Davidson:1981zd,Wilczek:1982rv,Davidson:1983fy,Davidson:1984ik,Berezhiani:1989fp}. 
 }. 
According to the FN mechanism, a $U(1)$ flavor symmetry which is denoted as  $U(1)_F$ is often used. 
If the $U(1)_F$ is a global symmetry, it may be violated by quantum gravity effects according to the standard lore
\cite{Misner:1957mt,Banks:1988yz,Abbott:1989jw,Coleman:1989zu,Garfinkle:1990qj,Horowitz:1991cd,Kallosh:1995hi,Polchinski:2003bq,Banks:2010zn,Harlow:2018jwu,Harlow:2018tng}.
If the $U(1)_F$ is gauged as in string motivated models, it can be also broken to a 
discrete symmetry
\cite{BerasaluceGonzalez:2011wy,Ibanez:2012wg,BerasaluceGonzalez:2012vb}
as a discrete gauge symmetry \cite{Krauss:1988zc,Alford:1989ch,Preskill:1990bm,Alford:1990mk,Alford:1990pt,Alford:1992yx}.
Flavor models with discrete symmetries are being well studied recently, and the vacuum analyses in such models appear to be complex
owing to a number of scalar fields \cite{Altarelli:2010gt,Ishimori:2010au,King:2013eh,King:2015aea}. See also, e.g., Refs.~\cite{Kobayashi:2006wq,Abe:2009vi} for string models~\footnote{
See, e.g.,  Refs.~\cite{Feruglio:2017spp,Kobayashi:2018vbk} for recent models with a flavor modular symmetry
and also Refs.~\cite{Hamidi:1986vh,Dixon:1986qv,Cremades:2003qj,Cremades:2004wa} for such interactions in string models.
}.
In this paper, we will focus on a simple model with a discrete abelian flavor symmetry.

In general, a flavon $S$ can always have couplings to the Higgs boson $H$, such as  
\begin{align}
\label{eq-SHcoup}
V \ni  c\abs{S}^2 \abs{H}^2 \quad \mathrm{or}\quad c' \frac{S^{N}}{\Lambda^{N-2}} \abs{H}^2.
\end{align}
The former will appear for a $U(1)_F$ flavor symmetry, whereas both two terms can appear for a $\Zn{N}$ flavor symmetry,
which is denoted as $\Zn{N}^F$ hereafter.
Once the flavon acquires VEV, 
the Higgs boson gets a mass of $c \vev{S}^2$ or $c' \eps^{N-2} \vev{S}^2$ at tree-level.
Therefore, $\vev{S}^2$ should be comparable to
the electroweak (EW) scale 
unless $c$ and $c'$ are extremely suppressed. 
Such tiny couplings of $c, ~c'$ will cause another hierarchy problem.
In addition, there is also the hierarchy problem due to quadratic divergences.  
These facts motivate us to consider a supersymmetric (SUSY) model of the FN mechanism with a TeV scale $\langle S \rangle$ 
leading to a light flavon. Hence, $\Lambda (= \langle S \rangle/\epsilon)$ results in a cutoff scale much lower than the Planck scale.

In this paper, we propose a SUSY extension of the SM 
with a flavon and a $\Zn{N}^F$ symmetry~\footnote{
Models with a combination of Higgs doublets $H_u H_d$ as a flavon 
are studied in Refs.~\cite{Babu:1999me,Giudice:2008uua,Bauer:2015kzy}.  
Models with discrete FN symmetries are recently discussed in Refs.~\cite{Abbas:2017vws,Abbas:2018lga}. 
}.  
Holomorphy of the superpotential constrains flavor structure on top of a $\Zn{N}^F$ symmetry.
The model becomes 
predictive, 
because couplings between $S$ and $H$ are related to Yukawa hierarchy
as in Eq.~\eqref{eq-SHcoup} 
on top of coupling relations within the Minimal Supersymmetric Standard Model (MSSM). 
A discrete symmetry $\Zn{N}^F$ allows a self-coupling of the flavon in the superpotential, $W\ni S^N/\Lambda^{N-3}$,
and the coupling stabilizes flavon.\footnote{
For a $U(1)_F$ gauge symmetry, D-term potential will stabilize the flavon VEV.
For a global $U(1)_F$ , flavon VEV may be stabilized by similar ways as in (fl)axion models.} 
This model also provides a solution for the $\mu$-problem in the MSSM.
The Higgsino mass term in the superpotential, the so-called $\mu$-parameter, 
is written by $W\ni ({S}^m/{\Lambda^{m-1}}) H_u H_d$.
Here $m$ is determined by charges of Higgs superfields, 
and the coupling is consistent with $\Zn{N}^F$.
This mechanism is similar to the Next-to-MSSM (NMSSM),
in which the $\mu$-parameter is explained by the singlet VEV~\footnote{
See for reviews~\cite{Ellwanger:2009dp,Maniatis:2009re}, 
and also Ref.~\cite{Kim:1983dt}.
}. 
Crucial differences from the typical NMSSM are that 
the singlet $S$ is charged under a flavor dependent $\Zn{N}^F$ symmetry with $N>3$,
whereas similar interactions can be found in string models in the presence of 
many scalar fields \cite{Lebedev:2009ag}.
In addition, a cutoff scale $\Lambda$ is much smaller than the Planck scale.

Since the flavor symmetry $\Zn{N}^F$ controls not only the SM fermion mass hierarchy
but also the Higgs sector in this model, the vacuum structure of the scalar potential needs to be checked.
The hierarchical structure in the Higgs potential can give significant effects to
the EW symmetry breaking~\footnote{
In this paper, we call the potential consisting of Higgs doublets and flavon as the Higgs potential.
}. 
Even in the $\Zn{3}$-invariant NMSSM, parameters in the potential should be chosen carefully 
to obtain the realistic EW symmetry breaking \cite{Kanehata:2011ei,Kobayashi:2012xv,Beuria:2016cdk}.
A coupling constant for a flavon self-coupling in a superpotential should be sizable, 
so that the quartic coupling $\sim \abs{S}^4$ stabilizes the Higgs potential while 
extra minimum deeper than the EW vacuum does not exist. 
For the $\Zn{N}^F$ symmetry with $N>3$,   
the Higgs potential will be more likely to develop extra minimum,   
since the corresponding self-coupling of the flavon is given by $\sim \abs{S^{(N-1)}}^{2}/\Lambda^{2(N-3)}$. 
Hence the potential becomes flatter. 
We will discuss conditions to prevent extra minimum deeper than the EW vacuum. 
In addition, the $\Zn{N}^F$ symmetry also controls the mass matrix 
of the Higgs boson and flavon whose mass scales are below the flavon VEV. 
We will discuss new physics related to the light flavons and the Higgs bosons.

An another aspect of this model we will discuss is dark matter (DM) candidate. 
In the presence of a certain discrete symmetry, such as R-parity, 
the Lightest SUSY Particle (LSP) is a good candidate for the DM.
In particular, masses of the Higgsinos are quite predictive in this model, 
because the masses depend on the hierarchical structure coming from the $\Zn{N}^F$.    
Note that the Higgs/Higgsino sector of this model can be regarded as a special case of 
two Higgs doublet model amended by adding a flavon field and a pair of Higgsinos
which is a candidate for the DM.
Finally, domain wall problem may exist in this model \cite{Zeldovich:1974uw,Vilenkin:1984ib,Ellis:1986mq,Abel:1995wk}.
This can be solved, e.g., when 
$S$ develops VEV and $\Zn{N}^F$ is broken during/before inflation 
owing to a Hubble-induced mass generated by a coupling of $S$ to the inflaton \cite{Chigusa:2018yua}.

This paper is organized as follows. 
The model is introduced in Section~\ref{sec-model}. 
We show conditions of charge assignments under the discrete flavor symmetry 
to obtain the realistic Yukawa texture without introducing non-abelian gauge anomalies.  
In section~\ref{sec-pheno}, we study phenomenology of this model.
We will focus on the vacuum structure, DM and phenomenology related to a light flavon.
Section~\ref{sec-concl} is devoted to a conclusion.  
Analytic formulas in the Higgs sector, possible K\"ahler potential corrections
and the values of Yukawa couplings in a benchmark point 
are shown in Appendices~\ref{App-anal}, \ref{KYukawa}
and~\ref{App-Bench}, respectively.

\section{Model}
\label{sec-model} 
In this section, we introduce an abelian flavor symmetry $\Zn{N}^F$ 
and a flavon field $S$ whose the charge is $1$ against $\Zn{N}^F$: under $\Zn{N}^F$, a flavon transforms as 
\begin{align}
S \to e^{2 \pi i/N}S.
\end{align} 
The VEV explains the fermion mass hierarchy by the Froggatt-Nielsen (FN) mechanism.  
This model has a cutoff scale $\Lambda$ much lower than the Planck scale, 
so that the hierarchy is explained by a ratio of a low scale flavon VEV to $\Lambda$~\footnote{ 
The FN mechanism via an inverse ratio $\Lambda/\vev{S}$ is recently proposed~\cite{Alonso:2018bcg,Smolkovic:2019jow}.  
}, 
\begin{align}
\epsilon := \langle S \rangle /\Lambda.
\end{align} 
In the model, such a small VEV is also related to the Higgs potential 
through a coupling e.g. $c \abs{S}^{2m} \abs{H}^2/\Lambda^{2m-2}\sim \epsilon^{2m}\Lambda^2 |H|^2$, which does not induce a large Higgs mass parameter
due to an $\epsilon$ suppression.
The SUSY is further introduced for several reasons. 
As in the MSSM, 
the SUSY model is free from the gauge hierarchy problem 
and the LSP becomes the good DM candidate. 
The SUSY is a well-motivated way to constrain a scalar sector. 
For instance, quartic couplings are related to gauge or Yukawa coupling constants, 
especially the Higgs quartic coupling in the MSSM is consistent 
with the 125 GeV Higgs boson mass. 
In addition, the fermion hierarchy can be explained 
by a discrete symmetry $\Zn{N}^F$ (with $N={\cal O}(1)$)
due to the holomorphy of the superpotential as shown later.  
As in the NMSSM, 
a VEV of the flavon generates the Higgsino mass term 
and the $\mu$-problem is solved.    
The flavon in this model couples to all the particles, 
such as SM fermions, Higgs bosons as well as the DM,  
and their textures are controlled by the flavor symmetry $\Zn{N}^F$. 
Further, we will show that anomalies between $\Zn{N}^F$ and the SM gauge group can constrain 
a coupling between the flavon and the Higgs sector.
Additional discrete symmetries 
are discussed for 
avoiding experimental constraints.

In this paper, the \Kahler\ potential is assumed to be the minimal one.
Even if there exist the higher dimensional operators in the \Kahler\ potential, 
they will not drastically change our results.
We discuss possible effects from these operators including kinetic term corrections
in Appendix~\ref{KYukawa}. 
Without loss of generality, 
the leading terms in the \Kahler\ potential can be the canonically normalized form, 
\begin{align}
 K = \sum_I \Phi^\dag_I e^{V_I} \Phi_I,     
\end{align}
where $\Phi_I$'s are any chiral superfields in this model, 
and $V_I$'s are certain combinations of the vector superfields against $\Phi_I$. 
In the following, we will introduce a superpotential in this gauge basis.

\subsection{Flavon-Higgs sector}
The $\Zn{N}^F$-invariant superpotential in the model is given by 
\begin{align}
 W_{\mathbbm{Z}_N} = 
\frac{c_N}{N \Lambda^{N-3}} S^N + \frac{c_m}{m \Lambda^{m-1}} S^m H_u H_d
+ W_\mathrm{fermion}, 
\end{align}
where $\Lambda$ is the cutoff scale in the model. 
Here, $H_u H_d = H_u^+ H_d^- - H_u^0 H_d^0$. 
$c_N$ and $c_m$ are $\order{1}$ coefficients. 
The integer $m$ is related to the charges of Higgs bosons, $n_{H_u}$ and $n_{H_d}$ 
as $m + n_{H_u} + n_{H_d} \equiv 0$ modulo $N$.  
The superpotential involving the SM fermions, 
$W_\mathrm{fermion}$, is introduced in the next subsection. 
Throughout this paper, 
we neglected the higher-order terms suppressed by $\Lambda^N$ 
such as $W \ni S^{2N}/\Lambda^{2N-3}$.

The scalar potential of the flavon and the neutral Higgs (against $U(1)_{\rm em}$) is given by 
\begin{align}
 V_0 :=&\ V_\mathrm{soft}+ V_\mathrm{F}+ V_\mathrm{D}, \\ 
V_\mathrm{soft} :=&\ m_S^2 \abs{S}^2 +m_{H_u}^2 \abs{H^0_u}^2+m_{H_d}^2 \abs{H^0_d}^2
                               + \left(A_S \frac{S^N}{N \Lambda^{N-3}} - 
                                            A_H\frac{S^m}{m \Lambda^{m-1}} H^0_u H^0_d 
                                         +  \mathrm{h.c.}\right), \\
V_\mathrm{F} :=&\ \abs{c_N \frac{S^{N-1}}{\Lambda^{N-3}}
                                         -c_m \frac{S^{m-1}}{\Lambda^{m-1}} H^0_u H^0_d }^2 
                                  + \left(\abs{H^0_u}^2 + \abs{H^0_d}^2 \right) 
                                       \abs{c_m \frac{S^{m}}{m \Lambda^{m-1}} }^2,            \\
V_\mathrm{D} :=&\ \frac{g^2}{2} \left(\abs{H^0_u}^2-\abs{H^0_d}^2 \right)^2, 
\end{align} 
where $ H^0_u, H^0_d$ are neutral components of the Higgs doublets $H_u, H_d$, respectively. 
$V_\mathrm{soft}$, $V_F$ and $V_D$ come from soft SUSY breaking terms, 
F-term potential of the superpotential and the D-term potential, respectively.    
The quartic coupling constant of the D-term is related to the gauge coupling constants 
as $g^2 = (g_1^2+g_2^2)/4$, 
where $g_1$ and $g_2$ are the gauge couplings  constants of $U(1)_Y$ and $SU(2)_L$. 
In this paper, the soft parameters are assumed to be real.

The scalar fields are expanded around their vacuum as,  
\begin{align}
S := v_s + \frac{1}{\sqrt{2}} \left(h_s + i a_s \right),
\quad  
H_u^0 := v_u + \frac{1}{\sqrt{2}} \left(h_u + i a_u \right), 
\quad 
H_d^0 := v_d + \frac{1}{\sqrt{2}} \left(h_d + i a_d \right),   
\end{align}
where $v_u^2 + v_d^2 = v_H^2 \sim 174$ GeV.  
Suppose that $v_s \gg v_H$, 
the VEV of flavon $S$ is approximately determined by the scalar potential   
\begin{align}
\label{eq-Vs}
 V_S = m_S^2 \abs{S}^2 + \abs{c_N \frac{S^{N-1}}{ \Lambda^{N-3}} }^2 
                            + \left(A_S \frac{S^N}{N \Lambda^{N-3}}  
                                         +  \mathrm{h.c.}\right).  
\end{align}
The flavon VEV satisfies  
\begin{align}
\label{eq-solvs}
 v_s^{N-2} 
 \sim  \frac{\Lambda^{N-3}}{2(N-1) \abs{c_N}^2} 
      \left[ - A_S  + \sqrt{A_S^2 -4 (N-1) \abs{c_N}^2 m_S^2 }\right].    
\end{align}
At the potential minimum,  mass eigenvalues for the CP-even and CP-odd flavons
are given by
\begin{align}
\label{eq-mhs}
 m_{h_s}^2 =&\ 
         2(N-1)(N-2) \left(c_N \eps^{N-3} v_s \right)^2  
            + A_S (N-2) \eps^{N-3} v_s  + \order{v_H^2},  \\ 
 m_{a_s}^2 =&\ - N   A_S \eps^{N-3} v_s  + \order{v_H^2}.   
\label{eq-mas}
\end{align}
Here, the soft mass terms are eliminated by the vacuum condition.
In this limit, the minimization conditions for the doublet Higgs bosons are similar 
to that for the MSSM, 
\begin{align}
\label{eq-EWmu}
\frac{1}{2} m_Z^2  =&\ - \abs{\mueff}^2 + \frac{m_{H_u}^2 \tan^2\beta - m_{H_d}^2}{1-\tan^2\beta} 
                             \sim   - \abs{\mueff}^2 - m_{H_u}^2  + \order{ \frac{m_{H_d}^2}{\tan^2\beta} },  \\
 \frac{2\mueff \Beff}{\sin 2\beta} =&\ 2 \abs{\mueff}^2 + m_{H_u}^2 + m_{H_d}^2 + \laeff^2 v_H^2,   
\label{eq-EWb}
 \end{align}
where 
$\tan\beta = v_u/v_d$,  $\laeff = c_m \eps^{m-1}$, 
$\mu_{\mathrm{eff}} = c_m/m \cdot \eps^{m-1} v_s$ and  
$B_\mathrm{eff} = A_H /c_m + c_N m\eps^{N-3} v_s$. 
The full Higgs mass matrices and the vacuum conditions are shown in Appendix~\ref{App-anal}.

The Higgsino mass matrix with decoupled gauginos is given by 
 \begin{align}
  \begin{pmatrix}
   \tilde{H}_d & \tilde{H}_u & \tilde{S}    
  \end{pmatrix}
\begin{pmatrix}
                     0 & - \mu_\mathrm{eff} &  - m \mu_{\mathrm{eff}} \cdot v_u/v_s   \\
-\mu_\mathrm{eff} & 0  & - m \mu_{\mathrm{eff}} \cdot v_d/v_s     \\ 
- m \mu_{\mathrm{eff}} \cdot v_u/v_s    &  -m \mu_{\mathrm{eff}} \cdot v_d/v_s    
&  m_{\tilde{S}}   \\  
\end{pmatrix} 
  \begin{pmatrix}
    \tilde{H}_d \\ \tilde{H}_u \\ \tilde{S} 
  \end{pmatrix},  
 \end{align} 
where $\tilde{S}$ is a SUSY partner of the flavon $S$ called flavino, and
\begin{align}
m_{\tilde{S}} := c_N (N-1) \eps^{N-3} v_s -c_m (m-1)\eps^{m-1} \frac{v_u v_d}{v_s} 
\end{align} 
is approximately a mass of the flavino.
The charged Higgsino mass is given by $\mu_\mathrm{eff}$.

\subsection{SM fermion mass and mixing} 
\label{sec-fmm}
The  $\Zn{N}^F$-invariant superpotential involving the SM fermions is given by 
\begin{align}
\label{eq-Wyuk}
 W_\mathrm{Fermion} 
=&\  
   c^u_{ij}\left(\frac{S}{\Lambda}\right)^{\eta^u_{ij}}
   \ol{u}_{R_i}  Q_{L_j} H_u 
+ c^d_{ij}\left(\frac{S}{\Lambda}\right)^{\eta^d_{ij}} 
\ol{d}_{R_i}  Q_{L_j} H_d  \notag \\ 
&+\  c^e_{ij} \left(\frac{S}{\Lambda}\right)^{\eta^e_{ij}} 
\ol{e}_{R_i}  L_{L_j} H_d 
+ c^n_{ij} \left(\frac{S}{\Lambda}\right)^{\eta^n_{ij}}   
\ol{N}_{R_i}  L_{L_j} H_u 
+\ \frac{1}{2}  M_{ij} \ol{N}_{R_i} \ol{N}_{R_j},       
\end{align}
where $i,j = 1,2,3$ run over the three generations. 
Here, we assume that the right-handed neutrinos have charge $N/2$,  
so that they have Majorana masses~\footnote{ 
$N$ should be an even number from this assumption.  
}.
A scale of the Majorana masses is, in general, at arbitrary scale, 
while this might be identified as the cutoff scale of this model $\Lambda$. 
Indeed, this happens at the benchmark point shown in Appendix~\ref{App-Bench}. 
The powers $\eta^f_{ij}$ ($f=u,d,e,n$) obey 
\begin{align}
- \eta^u_{ij} \equiv &\ n_{H_u} + n_{u_i} + n_{Q_j},&
- \eta^d_{ij} \equiv &\ n_{H_d} + n_{d_i} + n_{Q_j}, \\ 
- \eta^e_{ij} \equiv &\ n_{H_d} + n_{e_i} + n_{L_j},&
- \eta^n_{ij} \equiv &\ n_{H_u} + n_{n_i} + n_{L_j},  
\end{align}
modulo $N$. 
Hereafter, ``$\equiv$'' stands for modulo $N$ if it is not mentioned explicitly.  
$n_X$ is a charge of a field $X$ under the $\Zn{N}^F$ flavor symmetry. 
The Yukawa couplings to the Higgs bosons are induced after the flavon $S$ acquire a non-zero VEV, 
\begin{align}
\label{eq-yukH}
Y^f_{ij} = c^f_{ij} \left(\frac{v_s}{\Lambda}\right)^{\eta^f_{ij}},\quad 
f = u,d, e, n. 
\end{align}
Since the maximum power of the Yukawa hierarchy is $N-1$ under the $\Zn{N}^F$, 
it is assumed that the size of suppression factor is as small as the top to up quark mass ratio:
\begin{align}
\label{eq-epsN}
 \eps^{N-1} = \left(\frac{v_s}{\Lambda}\right)^{N-1} 
= \frac{m_u}{m_t} 
\sim 7.5 \times 10^{-6}. 
\end{align}
A combination $H_uH_d/\Lambda^2$ can couple to the Yukawa couplings, depending on the charge assignment,
but these will give negligible effects for $v_s \gg v_H$. 
It is noted that the holomorphy of superpotential is important 
to prevent couplings involving $S^\dag$ to fermions. Since a charge of $S^\dag$ is $N-1$,
once a coupling of $\Lcal\ni\left(S/\Lambda\right)^{\eta}{\cal O}_{\rm yukawa}$ is allowed by the discrete symmetry, 
$\Lcal\ni(S^\dag/\Lambda)^{N-\eta}{\cal O}_{\rm yukawa}$ is also allowed, 
where ${\cal O}_{\rm yukawa}$ is a Yukawa type operator. 
Thus, the maximal power becomes $N/2$ for even $N$ or $(N+1)/2$ for odd $N$ effectively 
if there is no holomorphy in the Yukawa couplings. 
Hence it is assumed in this paper that the SUSY should remain unbroken
below the cutoff scale $\Lambda$ 
in order to explain the fermion hierarchy with a small $N$.

So far, we assumed that the \Kahler\ potential is the canonical one and the Yukawa couplings originate only from the superpotential,
but let us discuss Yukawa couplings from higher dimensional \Kahler\ potential.
To justify our discussion, Yukawa couplings from \Kahler\ potential of $K \ni S^\dag {\cal O}_{\rm yukawa}/\Lambda^2$ should be suppressed,
when it is compared to a superpotential contribution  $W \ni (S/\Lambda)^{N-1}{\cal O}_{\rm yukawa}$. 
Since a VEV of F-term of the flavon is given by 
\begin{align}
 \vev{F_S^\dag} = - \vev{\frac{\partial W}{\partial S}} 
                          = - c_N \eps^{N-3} v_s^2 +  c_m \eps^{m-1} v_u v_d, 
\end{align}
the Yukawa couplings from the \Kahler\ potential is as small as 
the ones from the superpotential, 
$\vev{F_S^\dag}/\Lambda^2 \sim \epsilon^{N-1}$, where $\epsilon^{N-1}$ is comparable to a smallest Yukawa coupling. 
Thus the higher dimensional terms in the \Kahler\ potential
does not alter the texture of Yukawa matrices in the superpotential 
and only change $\order{1}$ coefficients per order. 
We absorb this effect in definitions of the $\order{1}$ coefficients $c_{ij}^f$ in the superpotential. 
See Appendix \ref{KYukawa} for more details of \Kahler\ potential.

\subsubsection{An example: $N=4$ case}
We will consider $N=4$ case as an example of the minimal extension of the 
$\Zn{3}$-invariant NMSSM,
and we have $W \ni S^4/\Lambda$. 
In this case, we obtain
\begin{align}
\eps \sim 0.02,
\end{align}
and it may be able to explain the hierarchy of the charged fermions
and the CKM elements involving the third generation.  
The Cabbibo angle $\sim 0.22 \sim \epsilon^{1/2}$ is regarded as an $\order{1}$ value
and explained with an $\order{1}$ tuning of holomorphic Yukawa coupling. 
The matrix element $V_{cb}$ 
is naturally addressed by $\eps \sim 0.02$ as shown below. 
Smaller $N$ makes harder to explain the hierarchical structure 
of the fermion masses and mixing at the same time.
For example, in the case of $N=3$, 
$\eps \sim 0.004 \ll \abs{V_{cb}}$ is expected 
from the top to up quark mass ratio,
but this will be too small to explain the other hierarchies.    
For $N \ge 6$ cases, $\eps \sim 0.22$, which is often considered, is allowed. 
The superpotential $W \ni S^N/\Lambda^{N-3}$ with a larger $N$ makes the scalar potential flatter along the $S$ direction.   
Hence, more careful parameter choice will be required to stabilize the flavon field 
and to realize the realistic EW vacuum. 
The stability of the EW vacuum is non-trivial even in the usual NMSSM with 
$N=3$, as discussed in Refs.~\cite{Kanehata:2011ei,Kobayashi:2012xv,Beuria:2016cdk}.

In this case, an ansatz of hierarchical structure of the Yukawa matrices, 
\begin{align}
\label{eq-nYsamp}
 Y_u \sim &\ 
\begin{pmatrix}
 \eps^{3} & \eps^{3} & \eps^{2} \\
 \eps & \eps & 1 \\
 \eps & \eps & 1 \\
\end{pmatrix},
\quad 
Y_d \sim 
\eps^k 
\begin{pmatrix}
 \eps^{2} & \eps^{2} & \eps \\
 \eps & \eps & 1 \\
 \eps & \eps & 1 \\
\end{pmatrix}, \notag \\ 
Y_e \sim &\ 
\eps^{k}
\begin{pmatrix}
 \eps^2 & \eps^2 & \eps^2 \\
 1 & 1 & 1 \\
 1 & 1 & 1 \\
\end{pmatrix},
\quad 
Y_n \sim 
\eps^\ell 
\begin{pmatrix}
 1 & 1 & 1 \\
 1 & 1 & 1 \\ 
 1 & 1 & 1 \\
\end{pmatrix},
\end{align} 
leads to the charged fermion mass hierarchy
\begin{align}
(m_u, m_c, m_t) \sim (\eps^3, \eps, 1),\quad 
(m_d, m_s, m_b)\sim \eps^{k}(\eps^2, \eps, 1),\quad 
(m_e, m_\mu, m_\tau) \sim \eps^{k}(\eps^2, 1, 1),
\end{align} 
and the CKM and PMNS matrices 
\begin{align}
 V_\mathrm{CKM}\sim 
\begin{pmatrix}
 1 & 1 & \eps \\ 
 1 & 1 & \eps \\ 
\eps & \eps & 1 
\end{pmatrix},
\quad 
  V_\mathrm{PMNS}\sim 
\begin{pmatrix}
 1 & 1 & 1 \\ 
 1 & 1 & 1  \\ 
1 & 1  & 1 
\end{pmatrix},    
 \end{align}
which are consistent with the observed values. 
Here, $k \leq 1$ and $\ell \leq 3$ for $N=4$.  
The integer $k$ is related to $\tan\beta$ as $\eps^k m_t/m_b \sim \tan\beta$. 
Hence, $k \le 1$ is required for $1 \lesssim \tan\beta \lesssim 50$, 
so that the Yukawa coupling constants are perturbatively small.    
The integer $\ell$ related to a scale of the Majorana mass $M_0$ as 
$m_\nu \sim \eps^{2\ell} v_u^2/M_0$ via the see-saw mechanism, where $m_\nu$ is a neutrino mass.   
To realize the above Yukawa matrices,
the conditions of charge assignment are given by    
\begin{align}
\label{eq-npat} 
 n_{Q_i}\equiv&\  (n_{Q_3}-1,n_{Q_3}-1,n_{Q_3}),\quad   
& n_{L_i} \equiv&\   (n_{L_3},n_{L_3},n_{L_3}), \notag \\ 
 n_{u_i} \equiv&\  (n_{u_3}-2,n_{u_3},n_{u_3}) ,\quad 
& n_{e_i} \equiv&\  (n_{e_3}-2,n_{e_3},n_{e_3}) , \notag \\ 
 n_{d_i} \equiv&\  (n_{d_3}-1,n_{d_3},n_{d_3}), \quad 
& n_{n_i} \equiv&\  (2,2,2),  
\end{align}
and $\Zn{4}^F$ invariant conditions in the Yukawa couplings are shown as
\begin{align}
\label{eq-third}
n_{H_u} + n_{Q_3} + n_{u_3} \equiv 0, 
\quad  
n_{H_d} + n_{Q_3} + n_{d_3}+k \equiv 0, 
\notag \\ 
\quad  
n_{Q_3} + n_{d_3} \equiv  n_{L_3} + n_{e_3},   \quad
\ell + n_{H_u} + n_{L_3}+2 \equiv 0,
\end{align}
modulo $N$. 
Here, all discrete charges of $n$'s are integers.
The condition Eq.~\eqref{eq-npat} is to explain the hierarchy between the flavors  
and Eq.~\eqref{eq-third} is to explain the third generation fermion masses. 
The neutrinos have universal charge under the flavor symmetry $\Zn{4}^F$, 
so that the large mixing in the PMNS matrix is realized. 
With this ansatz, $\order{1}$ values of PMNS matrix are naturally explained by the flavor symmetry $\Zn{4}^F$.
The small hierarchies between neutrino masses are explained
with a choice of ${\cal O}(1)$ Yukawa couplings 
since a neutrino mass squared is proportional to fourth power of the Yukawa couplings.
It is possible to explain the hierarchal neutrino masses
also by an introduction of additional discrete symmetries~\cite{Altarelli:2010gt,Ishimori:2010au,Ishimori:2012zz,Hernandez:2012ra,King:2013eh,King:2014nza,Tanimoto:2015nfa,King:2017guk,Petcov:2017ggy}. 
In this paper, we do not introduce such additional symmetries, 
and we show a set of values of $\order{1}$ coefficients 
which explain the neutrino mass differences accidentally.     
See numerical values exhibited in Appendix \ref{App-Bench} 
for realistic Yukawa couplings and Majorana masses.

\subsubsection{Constraint from anomalies in $N=4$ case}
We show that a way of a coupling of $S$ to the Higgs sector, $W \ni S^m H_u H_d/\Lambda^{m-1}$, is constrained in terms of 
anomalies of the discrete symmetry between $\Zn{4}^F$ and the SM gauge symmetries.
The abelian discrete symmetry $\Zn{4}^F$ may potentially induces anomalies~\cite{
Ibanez:1991pr,Ibanez:1991hv,Ibanez:1992ji,Dreiner:2005rd,Araki:2007zza,Araki:2008ek}. 
If there exist such anomalies, the discrete symmetry is no longer a symmetry in theories
and  explicit violation terms against the discrete symmetry are induced by quantum effects.
The anomalies of $\Zn{4}^F$- $SU(2)_L^2$ and $\Zn{4}^F$-$SU(3)_C^2$ are absent if  
\begin{align}
\label{eq-anom23}
 \Acal_{SU(2)_L} =&\  n_{H_u} + n_{H_d} + \sum_i \left(3 n_{Q_i} + n_{L_i} \right) 
                           \equiv n_{H_u}+n_{H_d} + n_{Q_3} + 3 n_{L_3} - 2  \equiv 0, \notag \\ 
 \Acal_{SU(3)_C} =&\  \sum_i \left(2 n_{Q_i} + n_{u_i} +n_{d_i}\right) 
                           \equiv 2 n_{Q_3}+3n_{u_3}+3n_{d_3} -3   \equiv 0, 
\end{align}
are satisfied.
Here, $4 n_{\Phi_I} \equiv 0$ modulo $4$, where $\Phi_I$'s are any chiral superfields. 
The conditions Eqs.~\eqref{eq-third} and \eqref{eq-anom23} are arranged to  
\begin{align}
\notag 
& n_{H_u} + n_{H_d} \equiv 3(1+k),\quad   n_{Q_3}+3 n_{L_3} \equiv 3 + k,  
\quad
n_{u_3} \equiv 3(n_{H_u} + n_{Q_3}),
\\ 
\label{eq-n3s}
& 
 n_{d_3} \equiv -k + 3 (n_{H_d}+n_{Q_3}),\quad   
 n_{e_3} \equiv 3 (1+n_{H_d}+n_{Q_3}),  \quad
2 - \ell \equiv n_{H_u} +n_{L_3}   .
\end{align}
From the above, $W \ni S^m H_u H_d/\Lambda^{m-1}$ shows the charge relation of 
\begin{align}
m + n_{H_u} + n_{H_d} = m+ 3(1+k) \equiv 0
\end{align}
modulo $4$, thus it is found that $m=1,2$ for $k=0,1$, respectively.  
Note that $m=3$ is not allowed.
Now there exist nine parameters of 
$(n_{Q_3}, n_{u_3}, n_{d_3},n_{L_3},n_{e_3},n_{H_u},n_{H_d}, k, \ell)$
and six constraints of Eq.~\eqref{eq-n3s}.
Altogether, we can regard $(k, n_{H_u}, n_{Q_3})$ as three free parameters, 
so there are $2\times 4\times 4 = 32$ ways to choose them.

Although from the bottom-up viewpoint we do not know a normalization of the $U(1)_Y$ 
which may be embedded into a larger (grand unified) gauge group at a higher energy scale~\footnote{
There will be no problem if $U(1)_Y$ is not embedded into a larger gauge group at high-energies, 
even if there is this type of anomaly.  
},
the anomaly of $\Zn{4}^F$ with $U(1)_Y$ could provide some insights. 
With a normalization factor $\mathcal{N}_Y$ which is assumed to be fractional,
the anomaly-free condition is given by   
\begin{align}
\Acal_{U(1)_Y} 
=&\  \mathcal{N}_{Y} \left[ 
      \frac{1}{2} (n_{H_u}+n_{H_d}) + \sum_i \left(\frac{1}{6} n_{Q_i}+\frac{4}{3}n_{u_i}+\frac{1}{3} n_{d_i}
              + \frac{1}{2} n_{L_i} + n_{e_i}  \right) \right]  \notag \\
=&\  \frac{\mathcal{N}_{Y}}{3} \left[ 
      \frac{3}{2} (n_{H_u}+n_{H_d}) + \frac{3}{2}n_{Q_3}+12n_{u_3} +3 n_{d_3} +\frac{9}{2} n_{L_3} 
        + 9 n_{e_3}  - 16 \right]  \equiv 0.   
\end{align}
Suppose $n_Y :=  \Ncal_Y/3 \in \mathbb{Z} $, the above equation is rewritten as
\begin{align}
\Acal_{U(1)_Y} \equiv  n_Y \left[ 
      \frac{3}{2} (n_{H_u}+n_{H_d}+n_{Q_3}+3n_{L_3} ) +3 n_{d_3}  +  n_{e_3}  \right]  
\equiv  n_Y (3 k + 2p ) \equiv 0,  
\end{align}
where the integer $p$ is defined through 
\begin{align}
n_{H_u}+n_{H_d}+n_{Q_3}+3n_{L_3} = 6 + 4 (k + p).   
\end{align}
Here, we used $n_{H_u}+n_{H_d} + n_{Q_3} + 3 n_{L_3} - 2  \equiv 0 $ 
from $\Zn{4}^F$-$SU(3)_C^2$ anomaly.
If $k = 0$ $(1)$, $3k+2p$ is even (odd). 
In particular, the anomaly-free condition is satisfied independent of $n_Y$ 
if $k=0~(m=1)$ and $p$ is even. 
On the other hand, the condition is trivial if $n_Y$ is a multiple of 4
and hence ${\cal N}_Y$ is a multiple of 12. 
For instance, $U(1)_Y$ might be embedded into a $U(12)$ theory in this case,
since ${\cal N}_Y$ is associated with a rank of a gauge group into which $U(1)_Y$ might be embedded.

The fermions in this model can also induce the gravitational anomaly~\cite{Araki:2008ek},  
\begin{align}
 \Acal^\mathrm{vis}_\mathrm{grav} = &\ 
      n_S+ 2(n_{H_u}+n_{H_d})+\sum_i \left(6n_{Q_i}+3n_{u_i}+3n_{d_i}+2n_{L_i}+n_{e_i}+ n_{n_i}   \right)  \notag \\
     \equiv&\ n_{u_3} + n_{d_3} + 3n_{e_3}. 
\end{align}
If there is no other particle charged under the $\Zn{4}^F$ symmetry,
the anomaly-free condition is given 
by $\Acal^\mathrm{vis}_\mathrm{grav} \equiv 0$ modulo $2$.   

\begin{table}[t]
\centering 
\caption{
\label{tab-chargek0}
Values of ($\ell$, $m$, $\tilde{\Acal}_{Y} $, $\Acal_\mathrm{gr}$) for $k=0~(m=1)$ 
with given $n_{H_u}$ and $n_{Q_3}$.  
The other charges are determined through Eqs.~\eqref{eq-npat} and~\eqref{eq-n3s},  
so that the hierarchy pattern Eq.~\eqref{eq-nYsamp} is realized and
the anomalies of  $\Zn{4}^F$ are vanishing in the SM non-abelian gauge groups. 
$\tilde{\Acal}_Y = \Acal_{U(1)_Y}/n_Y$ is the normalized anomaly of $\Zn{4}^F$ in the $U(1)_Y$ gauge group.  
If $n_Y$ is a multiple of 4, the anomaly is absent even for $\tilde{\Acal}_Y \equiv 2$.
We find $\ell = 0$ and $2$ for $\Acal_\mathrm{gr} \equiv 0$ modulo 2.
}
\newcolumntype{C}{>{\centering\arraybackslash}X}
 \begin{tabularx}{10cm}{CCC|CCCC}  \hline 
$k$&$n_{H_u}$&$n_{Q_3}$ & $\ell$& $m$ & $\tilde{\Acal}_{Y}$ & $\Acal_\mathrm{gr}$  \\ \hline\hline 
 0 & 0 & 0 & 1 & 1 & 0 & 1 \\
 0 & 0 & 1 & 0 & 1 & 2 & 0 \\
 0 & 0 & 2 & 3 & 1 & 0 & 1  \\
 0 & 0 & 3 & 2 & 1 & 2 & 0 \\ \hline 
 0 & 1 & 0 & 0 & 1 & 0 & 0  \\
 0 & 1 & 1 & 3 & 1 & 2 & 1  \\
 0 & 1 & 2 & 2 & 1 & 0 & 0   \\
 0 & 1 & 3 & 1 & 1 & 2 & 1   \\ \hline 
 0 & 2 & 0 & 3 & 1 & 0 & 1   \\ 
 0 & 2 & 1 & 2 & 1 & 2 & 0   \\ 
 0 & 2 & 2 & 1 & 1 & 0 & 1   \\ 
 0 & 2 & 3 & 0 & 1 & 2 & 0   \\ \hline 
 0 & 3 & 0 & 2 & 1 & 0 & 0   \\
 0 & 3 & 1 & 1 & 1 & 2 & 1   \\ 
 0 & 3 & 2 & 0 & 1 & 0 & 0   \\ 
 0 & 3 & 3 & 3 & 1 & 2 & 1   \\ \hline\hline 
 \end{tabularx}
\end{table}

\begin{table}[t]
\centering 
\caption{
\label{tab-chargek1}
The same figure as Table~\ref{tab-chargek0}, 
but $k=1~(m=2)$.
If $n_Y$ is a multiple of 4, the anomalies in the $U(1)_Y$ group is absent.
We find $\ell = 0$ and $2$ for $\Acal_\mathrm{gr} \equiv 0$ modulo 2.
}
\newcolumntype{C}{>{\centering\arraybackslash}X}
 \begin{tabularx}{10cm}{CCC|CCCC}  \hline 
$k$&$n_{H_u}$&$n_{Q_3}$ & $\ell$&$m$  & $\tilde{\Acal}_{Y}$ & $\Acal_\mathrm{gr}$  \\ \hline\hline
 1 & 0 & 0 & 2 & 2 & 1 & 0 \\
 1 & 0 & 1 & 1 & 2 & 3 & 1 \\
 1 & 0 & 2 & 0 & 2 & 1 & 0 \\
 1 & 0 & 3 & 3 & 2 & 3 & 1 \\ \hline 
 1 & 1 & 0 & 1 & 2 & 1 & 1 \\
 1 & 1 & 1 & 0 & 2 & 3 & 0 \\
 1 & 1 & 2 & 3 & 2 & 1 & 1 \\
 1 & 1 & 3 & 2 & 2 & 3 & 0 \\ \hline 
 1 & 2 & 0 & 0 & 2 & 1 & 0 \\
 1 & 2 & 1 & 3 & 2 & 3 & 1 \\
 1 & 2 & 2 & 2 & 2 & 1 & 0 \\
 1 & 2 & 3 & 1 & 2 & 3 & 1 \\ \hline 
 1 & 3 & 0 & 3 & 2 & 1 & 1 \\
 1 & 3 & 1 & 2 & 2 & 3 & 0 \\
 1 & 3 & 2 & 1 & 2 & 1 & 1 \\
 1 & 3 & 3 & 0 & 2 & 3 & 0 \\ \hline\hline 
 \end{tabularx}
\end{table}

Tables~\ref{tab-chargek0} and~\ref{tab-chargek1} show the patterns of powers $k, m, \ell$, 
the $U(1)_Y$ and gravitational anomalies 
when the realistic patterns of Yukawa couplings are realized 
and the $SU(2)_L$ and $SU(3)_c$ anomalies are absent. 
The lists for $k=0~(m=1)$ and $k=1~(m=2)$ are shown 
in Table~\ref{tab-chargek0} and~\ref{tab-chargek1}, respectively. 
The charges of the other chiral superfields 
are determined through Eqs.~\eqref{eq-npat} and~\eqref{eq-n3s}. 
Note that only $(k,\ell, m)$ are relevant to Yukawa couplings.
It is noted that only $\ell = 0,2$ are available for $\Acal_\mathrm{gr} \equiv 0$,
while $\tilde{\Acal_Y}=\Acal_{U(1)_Y}/n_Y$ is even (odd) for $k=0$ ($k=2$) as already stated. 
The $U(1)_Y$ anomaly is absent independent of $n_Y$ for $k=0$ case, if $\tilde{\Acal_Y} \equiv 0$ modulo 4. 
However, $k=0$ gives unstable Higgs potential as shown in the next section.  
For $k=2$, $\tilde{\Acal_Y}$ gives an odd number, therefore $n_Y$ should be a multiple of $4$, 
i.e., the $U(1)_Y$ normalization of ${\cal N}_Y$ should be a multiple of $12$.

\subsection{Particle stability and discrete symmetry}
Here, we discuss necessities of additional discrete symmetries, focusing on the proton decay.
The $\Zn{N}^F$ symmetry will not be enough to suppress unwanted higher-dimensional operators.  
Some combinations of baryon/lepton number violating operators 
are severely constrained by proton decay. 
The limits on the baryon/lepton number violating operators 
are~\cite{Barbier:2004ez,BenHamo:1994bq,Murayama:1994tc} 
\begin{align}
 \la_B\la_L  \lesssim \order{10^{-27}} ,\quad 
\kappa^{-1}  \gtrsim  \order{10^{27}\ \mathrm{GeV}} .  
\end{align}
Here, $\la_B$ and $\la_L$ are Yukawa couplings 
for the dimension-4 baryon number violating operator $\ol{u}_R\ol{u}_R\ol{d}_R$ 
and lepton number violating operator $L_LL_L\ol{e}_R, L_LQ_L\ol{d}_R$.  
$\kappa$ is a coupling constant for dimension-5 operators 
such as $Q_LQ_LQ_LL_L, \ol{u}_R\ol{u}_R\ol{d}_R\ol{e}_R$. 
These tiny coupling constants can not be explained by the $\Zn{4}^F$ symmetry.
Thus there should be additional symmetry to control these couplings.

There are several candidates which can forbid these operators. 
In the MSSM, the matter parity $M_2$, the so-called R-parity, 
is introduced for this purpose~\cite{Farrar:1978xj,Dimopoulos:1981dw,Dimopoulos:1981zb}.  
An important consequence of the R-parity is that the LSP becomes stable 
and it can be a good candidate for the dark matter (DM).  
The dimension-4 operators are forbidden by the R-parity, but the dimension-5 operators are not. 
Another candidate is known as the baryon triality, $B_3$, 
which prevents the baryon number violating operators 
while permits the lepton number violating ones~\cite{Ibanez:1991pr}. 
The baryon triality successfully ensures the proton stability, 
however, the LSP becomes unstable due to the lepton number violating interactions. 
The so-called proton hexiality $P_6$ prohibits 
all the baryon and lepton number violating dimension-4 and 5 operators, 
and the LSP is stable~\cite{Babu:2003qh,Dreiner:2005rd}. 
All of these three discrete symmetries are anomaly-free. 
The charge assignments under these discrete symmetries are listed in Table~\ref{tab-phex}. 
\begin{table}[t]
\centering
\caption{
\label{tab-phex}
Charges under R-parity $M_2$, baryon triality $B_3$ and proton hexiality $P_6$. 
}
 \begin{tabular}[t]{c|ccccccc} \hline 
        & $Q_L$ & $\ol{u}_R$& $\ol{d}_R$ & $L_L$ & $\ol{e}_R$ & $H_u$ & $H_d$ \\ \hline\hline
$M_2$ & 1 & 1 & 1 & 1 & 1 & 0 & 0 \\ 
$B_3$ & 0 & 2 & 1 & 2 & 2 & 1 & 2 \\ 
$P_6$  & 0 & 1 & 5 & 4 & 1 & 5 & 1 \\
\hline 
 \end{tabular}
\end{table}

A discrete R symmetry in SUSY theories is an interesting 
possibility~\cite{Kappl:2010yu,Lee:2010gv,Lee:2011dya,Dreiner:2013ala,Chen:2013dpa}. 
The anomaly-free $\Zn{M}^{R}$ symmetry 
prohibits the unwanted higher dimensional operators as well as the $\mu$-term, 
and stabilizes the LSP in the MSSM~\cite{Lee:2010gv,Lee:2011dya}. 
In this model, the flavon field $S$ must have non-zero charge under the discrete R-symmetry 
in order to write down the self-coupling $S^N$ in the superpotential.  
Note that a superpotential have a non-zero R-charge $2$ modulo $M$. 
This causes additional selection rules for the Yukawa couplings of the SM fermions.  
Hence, $N$ of the discrete symmetry $\Zn{N}^F$ 
may have to be so large that the SM fermion mass and mixing are explained.

We are interested in the simplest way to explain the observed fermion properties 
and the neutralino DM at the same time. 
For this purpose, 
we consider a model with 
$\Zn{4}^F \times M_2~{\rm or}~\Zn{4}^F \times P_6$. 
Phenomenology of the model with $M_2$ and that with $P_6$ are similar to each other, 
except for the presence of the proton decay. 
Models with $M_2$ will be excluded if
a cutoff scale of the dimension-5 operators is $\Lambda$ that is much smaller
than the conventional GUT scale or string/Planck scale.
Such dangerous operators are forbidden by imposing the proton hexiality $P_6$.
Since a cutoff scale and coefficients of such dimension-5 operators
depend on UV model-building,
also models with the R-parity $M_2$ may be allowed.
In the following, we do not consider the higher dimensional operators
violating lepton/baryon number.

A spontaneous breaking of $\Zn{4}^F$ could produce stable domain walls
which alters the history of the successful standard cosmology~\cite{Zeldovich:1974uw,Vilenkin:1984ib,Ellis:1986mq,Abel:1995wk}. 
There exist several solutions for it.
The Planck suppressed operators which break $\Zn{4}^F$ explicitly
can make domain walls unstable, while keeping the low-energy physics unchanged~\cite{Abel:1996cr,Panagiotakopoulos:1998yw,Panagiotakopoulos:1999ah,Panagiotakopoulos:2000wp,Dedes:2000jp}. 
In the presence of a negative Hubble induced mass term 
for the flavon $S$ (and/or Higgs fields) during/before the inflation,
domain walls will be produced then and inflated away, 
hence the problem is solved~\cite{Chigusa:2018yua}.
We assume that the domain wall problem is solved in our model by one of these effects.

\section{Phenomenology}
\label{sec-pheno}
We will study vacuum stability and phenomenology related to the flavons
when the hierarchy Eq.~\eqref{eq-nYsamp} is realized and the anomalies of non-abelian gauge symmetries are absent for $N=4$.  
In the following, we will discuss models with the superpotential $W \ni (S^m/\Lambda^{m-1}) H_u H_d$, where $m=1,2$.   
In our analysis, we study cases where squarks/sleptons are heavier than $\order{10}$ TeV. 
There may be various flavor violating processes induced by sfermions depending on the soft parameters, 
but this is beyond a scope of this paper.

\subsection{Vacuum stability} 
\label{sec-vacuum}
We will show the vacuum stability is related to the power of $m=1,2$.  
The EW minimum will be unstable 
if there exist extra minimum deeper than it. 
The both CP-even and CP-odd flavon get positive mass squared when
\begin{align}
\label{eq-condAs}
-6 c_N^2 \eps v_s \lesssim A_S \lesssim 0,  
\end{align}
where $\order{v_H^2}$ corrections are neglected. 
With this condition, 
the flavon VEV in Eq.~\eqref{eq-solvs}
requires  $m_S^2 \sim \order{\eps^2 v_s^2}$.  
It is noted that the Higgs potential is approximately given by the flavon potential 
of Eq.~\eqref{eq-Vs} since the flavon VEV is supposed to be 
much larger than the Higgs ones.
Thus, the depth of the EW minimum is approximately given by 
\begin{align}
V_{S,\mathrm{min}} \sim - \order{\eps^2 v_s^4}.   
\end{align}
We always choose a solution with $v_s > 0$ from the two minimum satisfying Eq.~\eqref{eq-solvs}.  
These features are independent of $m$.  
We discuss the stability of the EW vacuum for $m=1$ and $2$ separately.

\subsubsection{$m=1: ~W = S^4/\Lambda + S H_u H_d$} 
As discussed in Section~\ref{sec-fmm}, 
we have $k=0$ and $Y_b \sim 1$ in this case. Thus
a large $\tan\beta$ is required to explain the top to bottom quark mass ratio.  
As a result, $-m_{H_u}^2 \sim \abs{\mueff}^2 \sim v_s^2$ is required in the EW vacuum.    
The potential along the $H_u^0$ direction with $H_d^0 = S = 0$ is given by 
\begin{align}
 V_{H_u} = m_{H_u}^2 \abs{H_u^0}^2 + \frac{g^2}{2} \abs{H_u^0}^4.
\end{align}
This potential always has the minimum if $m_{H_u}^2 < 0$ as required to realize  the EW minimum 
in a large $\tan\beta$ regime. 
The depth of this minimum is given by 
\begin{align}
 V_{H_u, \mathrm{min}} = - \frac{\left(m_{H_u}^2 \right)^2 }{2 g^2} \sim - \order{{v_s^4}}  
                               \ll V_{S, \mathrm{min}}. 
\end{align} 
This minimum is deeper than the EW minimum by $\order{\eps^2}$.   
Thus the EW minimum is expected to be unstable for $m=1$.   
Hereafter, we do not consider this case and focus on the case with $m=2$.

\subsubsection{$m=2:~ W = S^4/\Lambda + S^2 H_u H_d/\Lambda$} 
Since we find $k=1$ and $Y_b \sim \epsilon$ in this case, 
we have $\tan \beta\sim\order{1}$. 
The EW vacuum condition requires
$-m_{H_u}^2 \sim \abs{\mueff}^2 \sim \eps^2 v_s^2$,  
and then 
\begin{align}
V_{H_u,\mathrm{min}} \sim - \order{\eps^4 v_s^4} \gg V_{S,\mathrm{min}}.   
\end{align} 
The minimum along the $H_u^0$-direction is shallower than the EW minimum by $\order{\eps^2}$. 
In addition, there may be deeper minimum along 
the so-called D-flat and/or F-flat directions~\cite{Kobayashi:2012xv}.  
We parametrize a direction $\phi$ in the Higgs potential as 
\begin{align}
\phi : =  H_d^0 = \alpha^{-1} H^0_u = \gamma^{-1} S.   
\end{align}   
The EW minimum is on a direction with $\alpha = v_u/v_d =  \tan\beta$ and $\gamma = v_s/v_d \gg 1$.  
The D-flat direction corresponds to $\alpha =1$ 
and the F-flat direction of $F_S^\dag = - \partial_S W =0$ is $\gamma^2 = \alpha c_m/c_N \sim \mathcal{O}(1)$. 
Thus, additional minimum may appear along directions with $\alpha, \gamma \sim \order{1}$.

The potential along the $\phi$ direction is given by 
\begin{align}
 V_\phi = m_\phi^2 \abs{\phi}^2 + \left(A_\phi \frac{\phi^4}{\Lambda}  + \mathrm{h.c.}  \right) 
 + \la_\phi \abs{\phi}^4 + \kappa_\phi \frac{\abs{\phi}^6}{\Lambda^2},  
\end{align}
where 
\begin{align}
 m_\phi^2 \equiv &\ m^2_{H_d} + \abs{\alpha}^2 m_{H_u}^2 + \abs{\gamma}^2 m_S^2 ,\quad 
 A_\phi \equiv  \frac{\gamma^2}{4} \left(\gamma^2 A_S - 2 \alpha A_H \right), \\ 
\la_\phi \equiv &\ \frac{g^2}{2} \left( \abs{\alpha}^2-1 \right)^2,\quad 
 \kappa_\phi \equiv  \abs{\gamma}^2 \left(\abs{c_N \gamma^2 -c_m \alpha}^2 
                                   + \frac{\abs{c_m}^2}{4} \abs{\gamma}^2 (1+\abs{\alpha}^2) \right).   
\end{align}
The couplings for $\abs{S}^4$ and $\abs{S}^6$ terms are always positive real 
and the potential is always bounded from below except for a direction $\alpha=1$ 
and $\gamma = 0$.  
Assuming all the parameters are real, the minimum of this potential is given by
\begin{align}
\label{eq-phisq}
\phi^2 = \frac{\Lambda^2}{3\kappa_\phi} \left[-\left(\la_\phi + 2 \frac{A_\phi}{\Lambda} \right) + 
\sqrt{\left(\la_\phi+\frac{2A_\phi}{\Lambda} \right)^2 -3\kappa_\phi \frac{m_\phi^2}{\Lambda^2}} 
 \right].  
\end{align}
This minimum is absent if the right-hand side is negative or complex.  
In general, minimums tend to be appear for small values of $\la_\phi$ and $\kappa_\phi$.
Note that $\la_\phi$ vanishes for $\alpha = 1$
and $\kappa_\phi$ vanishes for $\gamma = 0$.

For $\gamma = 0$, the scalar potential is independent of the flavon.
At least one minimum exists along the $H_u^0$ direction, 
and its depth is shallower than the EW vacuum due to the suppression by $\eps$
as already stated above. 
Along the D-flat direction with $\alpha = 1$, $\la_\phi$ is also vanishing.   
Then quadratic term should be positive, 
\begin{align}
 m_{\phi}^2 = m_{H_u}^2 + m^2_{H_d}  
              \sim \frac{\eps v_s}{\sin2\beta} \left[ A_H + \left(c_N-\frac{c_m}{4}\sin2\beta \right)2c_m \eps v_s\right]> 0. 
\end{align}
Here, we used Eq.~\eqref{eq-EWb}.
Thus $A_H \gtrsim \order{\eps v_s}$ is required.

For $\alpha = 1$ and $\gamma \neq 0$, 
only $\la_\phi$ vanishes while $\kappa_\phi\ne0$. 
Since a large positive $m_\phi^2$ will prevent an exotic minimum,   
let us parametrize $m_{H_d}^2 = c_d \cdot \eps v_s^2$.  
For simplicity, we assume $c_d \gg \eps$ here, 
and will discuss the validity of this assumption later. 
From the EW minimum condition Eq.~\eqref{eq-EWb}, 
we have a relation of 
$A_H \sim \sin2\beta\cdot m_{H_d}^2/(\eps v_s) = c_d \sin 2\beta\cdot  v_s$. 
Since $A_\phi \sim - \gamma^2 A_H /2 < 0$ with neglecting $-A_s \ll A_H$,  
the inside of the square root of Eq.~\eqref{eq-phisq} should be negative to prevent a minimum along this direction. 
This requirement leads to the upper bound on $c_d$:
\begin{align}
 4 A_\phi^2 - 3\kappa_\phi m_\phi^2 \sim 
\left( c_d \eps^{-1} \gamma^4 \sin^2 2\beta  - 3 \kappa_\phi \right)m_{H_d}^2 < 0.  
\end{align}
Note that $\kappa_{\phi}$ is minimized by $\gamma^2=c_N/c_m$ for $\alpha =1$.
Thus the upper bound on $c_d$ reads
\begin{align}
\label{eq-cdcd}
 c_d \lesssim&\  
                 \frac{c_m^2}{2}   \frac{3\eps}{\sin^22\beta}  
=  \frac{3c_m^2}{8} \left(1+\frac{2}{\tan^2\beta} + \frac{1}{\tan^4\beta}  \right)  
         \times  \eps \tan^2\beta.  
\end{align}
This translates to an upper bound on the CP-odd Higgs boson mass $m_A$
(see Section~\ref{Yukawa-mass} and Appendix~\ref{App-anal} for the definition)
through the condition for the realistic EW symmetry breaking.  
With Eq.~\eqref{eq-cdcd}, our assumption $c_d \gg \eps$ is satisfied 
for $\tan^2 \beta \gg 1$. 
Note that $\tan\beta \sim \order{1}$ 
is required to obtain the realistic top to bottom quark mass ratio for $m=2$. 
We will numerically study the scalar potential with $A_H \sim \order{\eps v_s}$.

\begin{figure}[t]
\centering 
\begin{minipage}[c]{0.48\hsize}
 \centering
\includegraphics[height=70mm]{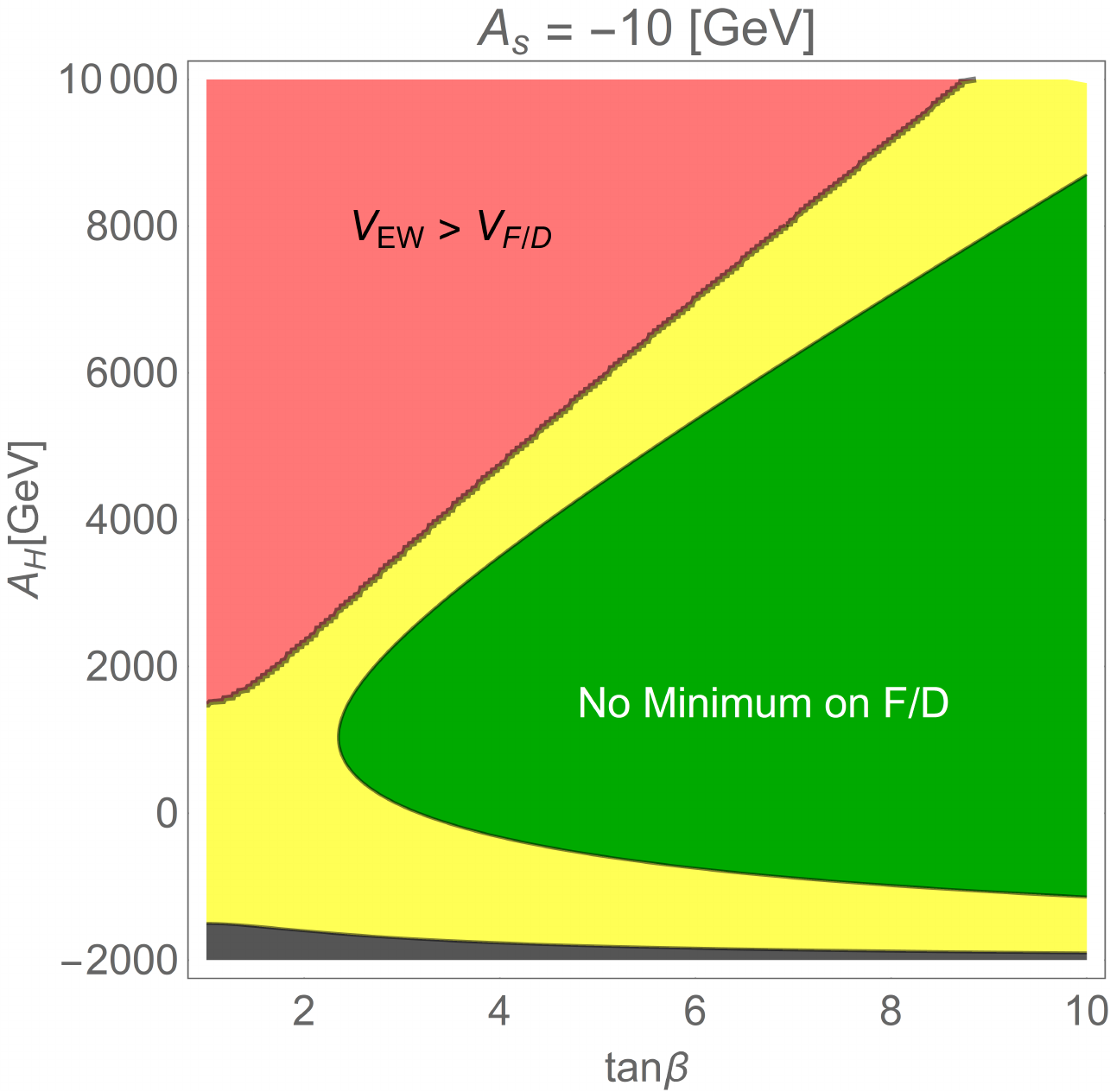} 
\end{minipage}
\begin{minipage}[c]{0.48\hsize}
 \centering
\includegraphics[height=70mm]{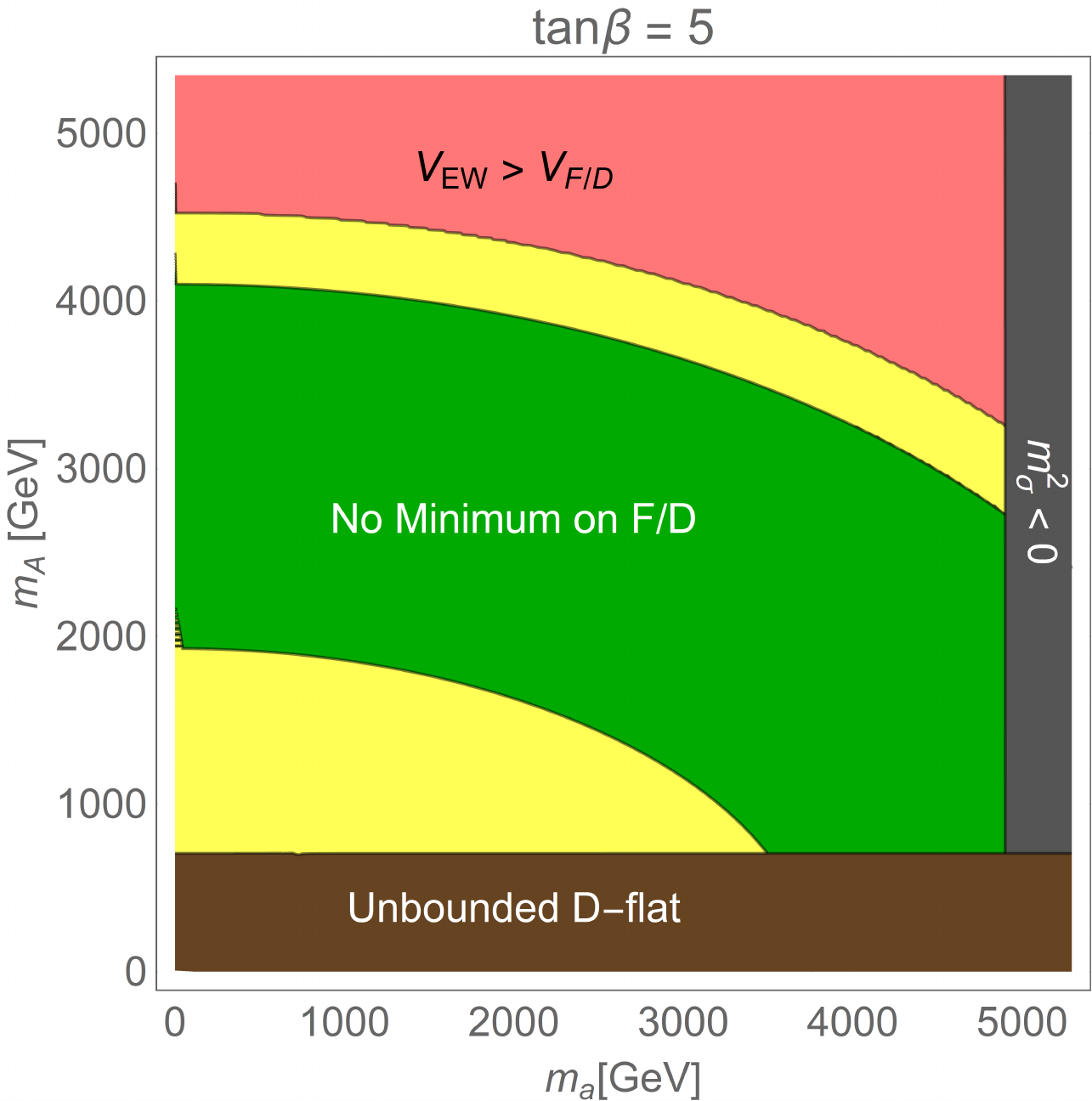} 
\end{minipage}
 \caption{
\label{fig-stability}
Parameter space where the EW minimum is deeper than the other minimum. 
$\eps v_s = 1.0$ TeV and $c_N = c_m = 1$. 
The green and yellow region have the stable EW minimum. 
} 
\end{figure}
Figure~\ref{fig-stability} shows the parameter space 
of $(\tan\beta,\ A_H)$ (left) and $(m_a,\ m_A)$ (right),
indicating a region where the EW minimum is deeper than the other vacuum. 
Here, the parameters are chosen to be $\eps v_s = 1.0$ TeV and $c_N = c_m= 1$. 
There is no minimum along the F/D-flat direction in green region.  
In the yellow region, the minimum exists along F/D-flat direction
but it is shallower than the EW minimum, while
in a red region the potential minimum along F/D-flat direction 
is deeper than the EW one.
The D-flat direction becomes unbounded from below in the brown region. 
The flavon mass becomes tachyonic, 
and then the point satisfying the EW condition is not a minimum in the gray region.  
Altogether, the green and yellow regions have the stable EW minimum.  
The wider parameter space is allowed with a larger $\tan\beta$, 
whereas the top to bottom quark mass ratio requires $\tan\beta$ to be $\order{1}$. 
In our analysis, we take $\tan\beta = 5$. 
In this case,  
the upper bound on the CP-odd Higgs boson is about $4$ TeV 
for $\eps v_s = 1$~TeV as shown in the right-panel.   
The limits on $A_H$ or $m_A$ will be relaxed for a larger $c_m$ due 
to a larger coupling of $\kappa_\phi$.

\subsection{Neutralino mass and dark matter physics} 
We will discuss DM physics under an assumption 
that the neutralino LSP is produced by the thermal freeze-out mechanism 
and they are not diluted after they are decoupled from the thermal bath. 
If stable LSP flavinos are produced thermally, 
they will be overproduced owing to a small cross section.
Thus, flavino LSP will not be considered in this paper.

First, let us consider cases in which the Higgsino is the 
LSP and is lighter than the flavino. 
For $N=4$ and $m=2$, the Higgsino and flavino masses are approximately given by 
\begin{align}
m_{\tilde{H}} \sim \mu_\mathrm{eff} \sim \frac{c_m}{2} \cdot \eps v_s ,\quad 
m_{\tilde{S}}   \sim 3 c_N  \cdot \eps v_s,  
\end{align}
where the mixing induced by the Higgs VEVs are neglected. 
Thus, $m_{\tilde{H}} \lesssim m_{\tilde{S}}$ can be realized 
when $c_m \lesssim 6 c_N$.  
The Higgsino can be identified as the DM particle as far as its mass is lighter than about 1.1 TeV, 
so that the LSP does not over-close the universe~\cite{Cirelli:2005uq,Cirelli:2007xd}.  
The Higgsino mass should be in a range, 
\begin{align}
  90\ \mathrm{GeV} \lesssim \mueff  \lesssim 1.1\ \mathrm{TeV},   
\end{align} 
where the lower bound comes from the LEP experiment~\cite{Jakobs:2001yg}.  
With assuming  $\mueff \sim 1$ TeV, the flavon VEV $v_s$ is expected to be $\order{100\ \mathrm{TeV}}$. 
The direct detection rate will be suppressed as far as the EW gauginos are 
much heavier than the Higgsino masses~\cite{Kawamura:2017amp}. 
This type of mass spectra, where Higgsinos are much lighter than other sparticles, 
 is the so-called natural SUSY. 
This would be obtained in Non-Universal Gaugino Mass (NUGM) scenario~\cite{Abe:2007kf,Bhattacharya:2009wv,Horton:2009ed,Gogoladze:2012yf,Brummer:2012zc,Yanagida:2013ah} 
or Non-Universal Higgs Mass scenario~\cite{Ellis:2002iu,Ellis:2002wv,Baer:2004fu,Baer:2005bu,Ellis:2008eu}. 
In particular,  the NUGM scenario with relatively heavy wino mass is interesting because 
the relatively large $m_{H_d}^2$ and small $m_{H_u}^2$ are realized simultaneously 
as a result of the renormalization group effects~\cite{Antusch:2012gv,Abe:2012xm}.    
This pattern of Higgs soft masses are consistent 
with the condition for the stable EW minimum discussed in the previous subsection.  
This feature was pointed out in the $\Zn{3}$ invariant NMSSM~\cite{Kawamura:2018qda}. 
The NUGM scenario is realized in GUT models~\cite{Ellis:1984bm,Ellis:1985jn,Drees:1985bx,Anderson:1996bg,Chakrabortty:2008zk,Martin:2009ad,Chakrabortty:2010xq,Kobayashi:2017fgl} 
as well as the so-called mirage mediation~\cite{Choi:2004sx,Choi:2005ge,Endo:2005uy,Choi:2005uz,Choi:2005hd,Kitano:2005wc}.  
The phenomenology of the mirage mediation in the NMSSM is discussed in Refs.~\cite{Asano:2012sv,Kobayashi:2012ee,Hagimoto:2015tua}.

Next, we shall consider cases in which the wino is the LSP for avoiding flavino overproduction.  
As the flavino mass can be comparable to the Higgsino mass,
the flavino is naturally heavier than $\order{100\ \mathrm{GeV}}$, 
so that the charged Higgsinos are heavier than the LEP bound.  
Even if the flavino is lighter than 100 GeV, 
its relic density can be lower than the observed value of DM only in restricted parameter space 
where the s-channel process is enhanced by the resonance or co-annihilation works due to degeneracies  
with some other particles~\cite{Djouadi:2008uj,Griest:1990kh}.  
A easier way to accommodate with the DM density may be 
that the wino lighter than about 2.7 TeV 
becomes the LSP~\cite{Hisano:2006nn,Cirelli:2007xd,Hryczuk:2010zi,Hryczuk:2011vi}.    
The hierarchy of the neutralinos are 
$\order{1\ \mathrm{TeV}} \sim M_2  < M_1 \ll  m_{\tilde{S}}, \mu_\mathrm{eff}$, 
so that the direct detection rate is suppressed by the heavy Higgsino mass of $\mu_{\rm eff} \gtrsim 10$ TeV. 
The flavon VEV is expected to be $\gtrsim \order{1\ \mathrm{PeV}}$ in this case.     
This type of mass spectrum, where gauginos are much lighter than other sparticles, 
is the so-called mini-split SUSY/pure gravity mediation 
scenario~\cite{Ibe:2011aa,Ibe:2012hu,Arvanitaki:2012ps,ArkaniHamed:2012gw}.
This spectrum of SUSY particles 
would be realized by the anomaly mediation~\cite{Randall:1998uk,Giudice:1998xp}  
in which gaugino masses are suppressed by the loop factor compared with the soft scalar masses.

\subsection{Yukawa interactions in mass basis}  
\label{Yukawa-mass}
We shall consider couplings of the scalars to the SM fermions
in cases with $k=1$ (hence $m=2$), in which $Y_{d,e} \propto \epsilon^{1}$ and $W \ni S^2 H_u H_d/\Lambda$.
In the gauge basis, the Higgs doublets are coupled to the SM fermions via the Yukawa couplings 
\begin{align}
 -\Lcal_{h_u, h_d} = \frac{h_u + ia_u}{\sqrt{2}} \ol{u}_R Y^u u_L
                          +\frac{h_d + ia_d}{\sqrt{2}}
                             \left( \ol{d}_R Y^d d_L + \ol{e}_R Y^e e_L \right) + \mathrm{h.c.},  
\end{align}
where the Yukawa matrices are defined in Eq.\eqref{eq-yukH}. 
The flavons are coupled to the SM fermions as 
\begin{align}
 -\Lcal_{h_s} = \frac{h_s + i a_s} {\sqrt{2}} 
                          \sum_{f=u,d,e} \ol{f}_R \Gamma^f f_L + \mathrm{h.c.}.  
\end{align}
These are  obtained by differentiating the usual Yukawa couplings with respect to $S$.
Hence the coupling matrices are given by 
\begin{align}
 \Gamma^f_{ij} = \eta^f_{ij} c^f_{ij}\left(\frac{v_s}{\Lambda}\right)^{\eta_{ij}^{f}-1} \frac{v_f}{\Lambda} 
                        = \frac{v_f}{v_s}\cdot \eta_{ij}^f Y^f_{ij},   
\end{align}
where $v_f = v_u$, $v_d$ for the up- and down-type fermions, respectively.  
Note that $\Gamma^u_{23}$ and $\Gamma^u_{33}$ are vanishing with
$\eta^u_{23} = \eta^u_{33} = 0$.
The flavon Yukawa couplings $\Gamma^f$ are more suppressed by $v_H/v_s$  
than those for the Higgs doublets. 
In addition, the flavino has Yukawa type interactions, 
\begin{align}
\label{eq-flavino}
-  \Lcal_{\tilde{S}} = 
        \sum_{f=u,d,e} \Gamma^f_{ij} \left[ \tilde{f}_{R_i} \tilde{S} f_{L_j} + \tilde{f}_{L_j} \ol{f}_{R_i} \tilde{S}\right]+ 
         \mathrm{h.c., }
\end{align}
where $\tilde{f}_{R_i}, \tilde{f}_{L_j}$ are sfermions.

We will rewrite these interactions in the mass basis.
The mass basis of the fermions, $\hat{f}_L, \hat{f}_R$ ($f=u,d,e$),  
are defined as 
\begin{align}
f_L = U^f_L \hat{f}_L,\quad f_R = U^f_R \hat{f}_R,\quad    
\left( U^f_R \right)^\dagger \left( Y^f_{ij} v_f \right)  U^f_L 
= \mathrm{diag} \left(m_{f_1}, m_{f_2}, m_{f_3} \right).   
\end{align} 
There are mixing between the Higgs bosons and flavon.  
The mass basis of the scalars are defined as 
\begin{align}
\begin{pmatrix}
 h_d \\ h_u \\ h_s 
\end{pmatrix}
=
R_S 
\begin{pmatrix}
 h \\ H \\ \sigma 
\end{pmatrix},
\quad 
 \begin{pmatrix}
 a_d \\ a_u \\ a_s 
\end{pmatrix}
=
R_P 
\begin{pmatrix}
 G \\ A \\ a 
\end{pmatrix},
\end{align}
where $h$ is the SM Higgs boson
and $G$ is a Nambu-Goldstone boson. 
The rotation matrices $R_S, R_P$ diagonalizes the Higgs mass matrices as 
\begin{align}
 R_S^T \mathcal{M}^2_S R_S = \mathrm{diag} (m_h, m_H, m_\sigma),
\quad  
 R_P^T \mathcal{M}^2_P R_P = \mathrm{diag} (0 , m_A, m_a). 
\end{align}
Here, $m_h $ is the SM Higgs boson mass.
A real scalar $\sigma$ ($a$) is defined as a scalar field in the mass basis 
whose a component of the rotation matrix $\left[R_{S}\right]_{3i}$ ($\left[R_{P}\right]_{3i}$) is the largest among the three scalars.
The scalar $\sigma$ $(a)$ is called as CP-even (CP-odd) flavon. 
The scalar mass matrices are shown in Appendix~\ref{App-anal}.

The Yukawa matrices $\hat{Y}^f$, $\hat{\Gamma}^f$ 
are defined in the mass basis of fermions, 
\begin{align}
 \hat{Y}^f   = \left(U^f_R\right)^\dagger Y^f U^f_L,
\quad  
\hat{\Gamma}^f= \left(U^f_R\right)^\dagger \Gamma^f U^f_L. 
\end{align}
The Higgs Yukawa coupling $\hat{Y}^f$ is diagonalized in this basis, 
but flavon Yukawa coupling $\hat{\Gamma}^f$ is not. 
The latter have the following textures: 
\begin{align}
 \hat{\Gamma}^u \sim  \frac{v_u}{v_s} 
\begin{pmatrix}
 \eps^3 & \eps^3 & \eps^2 \\ 
\eps^3   & \eps   & \eps^2 \\ 
\eps   & \eps   & \eps^2 
\end{pmatrix},
\quad 
 \hat{\Gamma}^d \sim  \frac{v_d}{v_s}
\begin{pmatrix}
 \eps^3 & \eps^3 & \eps^2   \\ 
\eps^4 & \eps^2 & \eps^3 \\ 
\eps^2 & \eps^2 & \eps   
\end{pmatrix},
\quad 
 \hat{\Gamma}^e \sim \frac{v_d}{v_s} 
\begin{pmatrix}
\eps^3 & \eps^3 & \eps^3 \\
 \eps^5 & \eps & \eps^5  \\ 
 \eps^5 & \eps^5 & \eps  \\ 
\end{pmatrix}.   
\end{align} 
Since $\Gamma^u_{23}=\Gamma^u_{33}=0$, these elements in $\hat{\Gamma}^u$ 
are obtained via mixing matrix for diagonalization. 
For $v_s \sim \order{10\ \mathrm{TeV}}$ realizing the heavy charged Higgsinos,  
$\hat{\Gamma}^f$ couplings are at most $\order{10^{-4}}$.  
Some off-diagonal elements are more suppressed 
than the Higgs Yukawa coupling in the gauge basis,  
since the some parts of the flavon couplings are aligned with the Higgs couplings,  
especially for the lower two rows in the Yukawa matrix of the charged leptons. 
This feature extremely suppresses the lepton flavor violation processes~\footnote{
The result is not changed by sub-leading terms suppressed by $(S/\Lambda)^{5}$
which may potentially exist in the superpotential.
}.

We finally write the Yukawa interactions in the mass basis between the SM fermions and scalars as  
\begin{align}
 - \mathcal{L}_\mathrm{yuk} 
=  \frac{1}{\sqrt{2}}
       \sum_{f=u,d,e} 
        \ol{\hat{f}}_R 
        \left[ h \ \hat{y}^f+\sigma\hat{\la}^{f,\sigma}+ a\hat{\la}^{f,a}  \right]
        \hat{f}_L + \mathrm{h.c.}.   
\end{align}
The Yukawa coupling of $h$ is given by 
\begin{align}
\label{eq-yHiggs}
 \hat{y}^f_{ij} 
= 
 \left[R_S\right]_{b1} \hat{Y}^f_{ij} 
  + \left[R_S\right]_{31} \hat{\Gamma}^f_{ij} 
            \sim 
                  \hat{Y}^f_{ij} 
                \left[1+ \order{\frac{v_f^2}{v_s^2}} \right], 
\end{align}  
where we take $b = 1$ for $f=d,e$ 
and $b = 2$ for $f=u$.
We have used $\left[R_S\right]_{3b} \lesssim \order{v_f/v_s}$ 
due to the hierarchical structure of the Higgs boson matrix as shown in Appendix~\ref{App-anal}. 
Those of the CP-even and CP-odd flavon couplings are given by 
\begin{align}
 \hat{\la}^{f,\sigma}_{ij} = 
                                           \left[R_S\right]_{b3} \hat{Y}^f_{ij} 
                                    + \left[R_S\right]_{33} \hat{\Gamma}_{ij}^f, \quad  
 \hat{\la}^{f, a}_{ij} = 
i \left(\left[R_P\right]_{b3} \hat{Y}^f_{ij} 
                                   + \left[R_P\right]_{33} \hat{\Gamma}_{ij}^f\right).
\end{align}
Flavor violating couplings of the SM Higgs boson is strongly suppressed by 
$v_f^2/v_s^2\lesssim 10^{-4}$ and will be negligible since $\hat{Y}^f$ is diagonal. 
The flavor violating couplings of 
$H, A$ and $H^\pm$
are also expected to be tiny similarly to $h$. 
With respect to $\sigma$ and $a$, both terms proportional to $\hat{\Gamma}^f$ and  $\hat{Y}^f$ 
contribute to the $\order{v_H/v_s}$ couplings, but only $\Gamma^f$ have non-zero off-diagonal elements in the mass basis. 
Altogether, we find
\begin{align}
\label{eq-flavonY}
 \hat{\la}^{u, \scal} \sim \rho_u \frac{v_u}{v_s}
\begin{pmatrix}
 \eps^3 & \eps^3 & \eps^2 \\ 
\eps^3   & \eps   & \eps^2 \\ 
\eps   & \eps   &   1 
\end{pmatrix},
\quad 
 \hat{\la}^{d, \scal} \sim \rho_d \frac{v_d}{v_s}
\begin{pmatrix}
 \eps^3 & \eps^3 & \eps^2   \\ 
\eps^4 & \eps^2 & \eps^3 \\ 
\eps^2 & \eps^2 & \eps   
\end{pmatrix},
\quad 
 \hat{\la}^{e,\scal} \sim  \rho_e \frac{v_d}{v_s}
\begin{pmatrix}
\eps^3 & \eps^3 & \eps^3 \\
 \eps^5 & \eps & \eps^5  \\ 
 \eps^5 & \eps^5 & \eps  \\ 
\end{pmatrix},    
\end{align}
where $\rho_f$ ($f=u,d,e$) are $\order{1}$ coefficients
and $\scal$ denotes both $\sigma$ and $a$. 
It is noted that $a t\ol{t}$ coupling is more suppressed by $1/\tan^2 \beta$
against that of $\sigma t\ol{t}$ owing to the difference of scalar mixing matrix.
See Appendix~\ref{App-anal} for more detailed discussions for the scalar mixing.

\subsection{Higgs physics}
We shall discuss Higgs decay modes.
Note that a small $\tan \beta$ is required 
from the anomaly-free charge assignment and the vacuum stability in this case.
The SM Higgs boson mass matrix has similar structure as in the MSSM, 
since the mixing with the flavon is suppressed by $\eps$. 
The contribution from mixing with the flavon to the SM-like Higgs boson mass squared is 
estimated as $\eps^2 v_H^2 \sim 10$ GeV$^2$.  
Hence, the effect is less than $\order{0.1\ \%}$ of the that from the D-term potential  
$\sim m_Z^2 \cos^2 2\beta$ and does not give significant effects.  
Depending on $\tan\beta$,  
the top squark mass has an upper bound to be consistent 
with the 125 GeV Higgs boson mass.
The upper bound is typically $100\ (10^4)\ \mathrm{TeV}$ 
for $\tan\beta = 4\ (2)$~\cite{Ibanez:2013gf}.  
This upper bound becomes tighter if there is a sizable mixing between top squarks. 
This upper bound is consistent with a typical value of the soft mass of the down-type Higgs boson,  
$m_{H_d}^2 \lesssim \tan^2 \beta \cdot \mu_{\rm eff}^2 \sim \tan^2 \beta \cdot (\eps v_s)^2 \sim \order{ \left(10\ \mathrm{TeV}\right)^2 }$.  
In our numerical analysis, 
we add a typical size of loop corrections, $\Delta m_{22} = (90~{\rm GeV})^2$, 
to $\Mcal^2_{S,22}$, which is the coefficient of $h_u^2$ in the scalar potential (see Appendix~\ref{App-anal}), 
by hand in order to explain $m_h\sim125$ GeV.
This does not give significant effects to phenomenology other than the Higgs boson mass 
itself due to the $\eps$ suppressed mixing.

The SM Higgs boson can decay to a pair of the CP-odd flavons if $2 m_a < m_h$. 
The relevant trilinear coupling between the SM Higgs boson and CP-odd flavons is given by 
\begin{align}
A_{h a a} \sim 
\frac{v_H}{\sqrt{2}}  \left(\eps^2 c_m^2 +  \frac{A_H}{\Lambda} \sin2\beta \right)
\sim \order{ \eps^2 v_H }. 
\end{align}
Neglecting the flavon mass, the branching fraction is given by 
\begin{align}
\mathrm{Br}\left(h\to aa\right) 
\sim \frac{\abs{A_{haa}}^2}{32\pi m_h \Gamma_h} 
  \sim  10^{-4} \times \left(\frac{A_{haa}}{0.07\ \mathrm{GeV}}\right)^2,   
\end{align} 
where $\Gamma_h$ is the decay width of the SM Higgs.
As discussed later, the CP-odd flavon with $m_a \lesssim m_t$ decays 
to $bb$ and $\tau\tau$ with about 80 $\%$ and 20 $\%$ branching fractions, respectively.  
We may have $4b$ and/or $2b2\tau$ signals from the Higgs boson decays, 
but these are much smaller than the experimental sensitivity~\cite{Carena:2007jk}.

As discussed in Section~\ref{sec-vacuum}, 
the CP-odd Higgs boson should be lighter than about 4 TeV for cases in which 
the Higgsino is the LSP with $m_{\tilde{H}} \sim \eps v_s\lesssim 1.1$ TeV,
so that the EW vacuum is stable.
Since the Higgs sector is similar to the MSSM, 
the CP-even Higgs $H$ and charged Higgs $H^\pm$ 
have almost same masses as the CP-odd Higgs boson $A$. 
The dominant decay mode of the neutral Higgs bosons, namely $H$ and $A$, 
will be a pair of top quarks 
because $\tan\beta \sim \order{1}$ is required.
That of the charged Higgs $H^\pm$ is a top quark and a bottom quark. 
In other words, the branching fractions to the leptonic modes, 
which are more strongly constrained~\cite{
Aaboud:2017sjh,Sirunyan:2018zut,Sirunyan:2019hkq,Aaboud:2018gjj},   
are suppressed owing to a small $\tan\beta$.  
There are substantial limits from the current searches for heavy Higgs bosons decaying to top quark at the LHC
only if $\tan\beta \lesssim 1$ 
and the top Yukawa coupling is enhanced~\cite{
Aaboud:2017hnm,Sirunyan:2019wph,Sirunyan:2019arl}.

\subsection{Flavon physics}
In this subsection, we shall discuss flavor violations mediated only by the flavons of $\sigma$ and $a$, and their decay modes.
The effects from the other particles will be enough suppressed 
if their masses are heavier than $\order{10\ \mathrm{TeV}}$. 
This model is more predictive than conventional flavon models 
due to the direct correlation between the Higgs potential and DM physics if Higgsino is the LSP.
The flavon VEV controls not only the Yukawa hierarchies which include flavon couplings to the fermions 
but also the Higgs mixing to the flavon and DM mass.
Hence, the VEV can be determined by DM physics.
Phenomenology of light flavon is discussed in Refs.~\cite{
Bauer:2016rxs,  
Tsumura:2009yf,Berger:2014gga,Diaz-Cruz:2014pla,Arroyo-Urena:2018mvl, 
Dorsner:2002wi,Huitu:2016pwk 
}. 
In general, the light flavons are accessible in flavor violating processes, 
such as $K$-$\ol{K}$ mixing,  $\mu \to e\gamma$ and $\mu \to e$ conversion~\cite{Bauer:2016rxs}.  
The top physics is also relevant because of its large Yukawa coupling.  
In particular, a sizable flavon coupling to $tc$ is predicted as ${\cal O}(v_u \epsilon/v_s) $
and may provide good signals 
at collider experiments~\cite{Bauer:2016rxs,Tsumura:2009yf,Arroyo-Urena:2018mvl}.   
Significant differences from the ordinal FN mechanism is that 
$\eps$ is assumed to be about $10^{-2}$ which is smaller by one order of magnitude 
than the usual value $\sim0.2$. 
In addition, some flavor violating couplings of the flavons, especially to charged leptons, 
are suppressed by the alignment with the Higgs Yukawa couplings.

\subsubsection{Lepton flavor violation} 
We will focus on flavor violating processes in the lepton sector.
The branching fraction of Lepton Flavor Violation (LFV) decays of
$\ell_i \to \ell_j \gamma$ is given by~\cite{Lavoura:2003xp},  
\begin{align}
\mathrm{Br}\left({\ell_i}\to {\ell_j \gamma}\right) =  
\frac{\alpha_e}{1024\pi^4 \Gamma_{\ell_i}} 
\left(m_{\ell_i}-\frac{m_{\ell_j}^2}{m_{\ell_i}}\right)^3 
 \left( \abs{\sigma_L}^2 + \abs{\sigma_R}^2 \right),      
\end{align}
where
\begin{align}
\sigma_L \simeq &\ \sum_{k=1,2,3}  \sum_{\scal=\sigma, a} \frac{1}{4 m_\scal^2} \left[ 
                    \left(m_{\ell_i} \hla^{e,\scal}_{jk} \hat{\la}^{e,\scal*}_{ik}+m_{\ell_j} \hla^{e,\scal *}_{kj}\hla^{e,\scal}_{ki} \right) 
                        F\left(\frac{m_{\ell_k}^2}{m_\scal^2} \right) 
           - m_{\ell_k}  \hla^{e,\scal}_{jk} \hla^{e,\scal}_{ki}   G\left(\frac{m_{\ell_k}^2}{m_\scal^2}  \right) 
             \right], 
\end{align}
and the loop functions are given by 
\begin{align}
 F(y) =&\ -\frac{y^3-6y^2+3y+6y \ln{(y)} + 2}{6(1-y^4)^4}, \quad 
 G(y) = \frac{y^2-4y+2\ln{(y)}+3}{(1-y)^3}.  
\end{align}
It is noted that $\sigma_R$ is obtained by formally replacing $\la^e_{ij} \to \la^{e*}_{ji}$. 
Here, $\Gamma_{\ell_i}$ is a width of lepton $\ell_i$. 
As $m_\sigma \gtrsim m_a$, with neglecting contributions from the CP-even flavon we estimate
\begin{align}
\mathrm{Br}\left(\mu\to e\gamma\right) 
\sim&\ \frac{\alpha_em_\mu^3}{1024\pi^4\Gamma_\mu} 
                 \frac{m^2_\mu}{16 m_a^4} 
        \abs{\hat{\la}^{e,a}_{12} \hat{\la}^{e,a}_{22}}^2  F\left(\frac{m_\mu^2}{m_a^2}\right)^2 \\ \notag 
\sim&\ 3 \times10^{-27} \times \left(\frac{10\ \mathrm{TeV}}{v_s}\right)^4
                                           \left(\frac{100\ \mathrm{GeV}}{m_a}\right)^4
                                           \left(\frac{\eps}{0.02}\right)^{8},  
\end{align}
where $\order{1}$ factors $\rho_e$ are simply replaced by unity.  
Note that the contributions enhanced by the tau lepton mass is more suppressed by powers of $\eps$
owing to the alignment. 
Thus the $\mu\to e\gamma$ is extremely suppressed by the higher powers of $\eps$ 
and is far below the experimental sensitivity even if the flavon is $\order{10\ \mathrm{GeV}}$. 
The other LFV decays, including three body decays like $\mu \to e\ol{e}e$, are also suppressed.

Let us give a comment about contributions from the flavino. 
For a simplicity, suppose that the soft parameters respect the fermion flavor structure and the sfermions are aligned with the fermions. 
Then the flavino couplings in the mass basis are also given by $\hat{\Gamma}^f$ in Eq.~\eqref{eq-flavino}. 
The largest contribution to $\mu\to e\gamma$ will come from the chirality enhanced effects  
which are proportional to the flavino mass if the corresponding sleptons have sizable left-right mixing. 
The contribution is roughly given by replacing $m_\mu^2/m_a^4 \to m_{\tilde{S}}^2/m^4_{\tilde{\ell}}$,  
and the ratio to the CP-odd flavon effect  is estimated as 
\begin{align}
\left(\frac{m_{\tilde{S}}^2}{m_{\tilde{\ell}}^4}\right)  \left(\frac{m_\mu^2}{m_a^4}\right)^{-1} 
 \sim 14 \times \left( \frac{m_{\tilde{S}}}{1\ \mathrm{TeV} }\right)^2  
                        \left( \frac{m_{a}}{100\ \mathrm{GeV} }\right)^4 
                        \left( \frac{5\ \mathrm{TeV} }{m_{\tilde{\ell}}}\right)^4.   
\end{align}
Thus the sparticle contributions are also far below the detectable level
when only the Yukawa couplings $\hat{\Gamma}^e$ cause flavor violation.

The $\mu$-$e$ conversion process in nuclei induced by flavons might be detectable~\cite{Bauer:2016rxs}.  
The conversion rate is given by~\cite{Kitano:2002mt,Cirigliano:2009bz},   
\begin{align}
\Gamma_\mathrm{conv} = 
4 m_\mu^5 \abs{m_p \tilde{C}^p_{SR} S^p + m_n \tilde{C}^n_{SR} S^n}^2 + \left(L\leftrightarrow R\right),  
 \end{align}
where $p$ and $n$ denote a proton and a neutron respectively, and
\begin{align}
 \tilde{C}^{p}_{SR}=&\ \sum_{q=u,d,s} C_{SR}^q f^p_{S_q} 
         + \frac{2}{27}\left(1-\sum_{q=u,d,s} f^p_{S_q}\right)  \sum_{Q=c,b,t} C_{SR}^Q,  \\ 
 \tilde{C}^{p}_{SL}=&\ \sum_{q=u,d,s} C_{SL}^q f^p_{S_q} 
         + \frac{2}{27}\left(1-\sum_{q=u,d,s} f^p_{S_q}\right)  \sum_{Q=c,b,t} C_{SL}^Q.   
\end{align}
Those for neutron are obtained by formally replacing $p\to n$. 
Here, only the scalar interactions are considered 
since the one-loop corrections to the dipole operator will be negligibly small
as deduced from discussions in $\mu\to e\gamma$.    
In this model, the coefficients $C_{SR}^q$ is given by 
\begin{align}
m_{q_i} C_{SR}^{q_i} = \sum_{\scal=\sigma, a} \frac{\hat{\la}^{e,\scal}_{12} }{2m_\scal^2} 
                                            \cdot  \mathrm{Re} \left(\hat{\la}^{q,\scal}_{ii} \right), 
\quad 
m_{q_i} C_{SL}^{q_i} = \sum_{\scal=\sigma, a} \frac{\hat{\la}^{e,\scal*}_{21} }{2m_\scal^2} 
                                            \cdot  \mathrm{Re} \left(\hat{\la}^{q,\scal}_{ii} \right).  
\end{align}
Following Ref.~\cite{Bauer:2016rxs}, 
we used the values for scalar form factors $f_{S_q}^{p,n}$ calculated 
in Refs.~\cite{Crivellin:2013ipa,Crivellin:2014cta} based on the lattice result~\cite{Junnarkar:2013ac}
and the overlap integrals $S^{p,n}$~\cite{Kitano:2002mt}.  
We compare with the current limit~\cite{Bertl:2006up} at SINDRUM II experiment for a gold target,  
and future limit at the DeeMe~\cite{Natori:2014yba}, COMET~\cite{Kuno:2013mha} 
and Mu2e~\cite{Abrams:2012er} experiments for an aluminum target, 
\begin{align}
 \mathrm{Br} \left(\mu \to e\right)^\mathrm{Au (Al)} = \frac{\Gamma_\mathrm{conv}}{\Gamma_\mathrm{capt}} 
 < 7\times 10^{-13}\ (6\times 10^{-17}), 
\end{align}
where $\Gamma_\mathrm{capt} = 13.07$ and $0.7054$ $[\times 10^{6} \cdot s^{-1}]$ 
in gold and aluminum~\cite{Kitano:2002mt,Suzuki:1987jf}, respectively.  
We will show the current and expected limits from $\mu\to e$ conversion 
in Figs.~\ref{fig-rslt1} and \ref{fig-rslt5} as below.

\subsubsection{Quark flavor violation}
\begin{table}[t]
\centering
\caption{\label{tab-valOs}
Numerical values of the hadronic matrix elements at $\mu = $ 1 TeV~\cite{Kawamura:2019rth,Kawamura:2019hxp}. 
}
\begin{tabular}{c|cccc} \hline 
                &   $\Ocal^{LR}_1(\mu)$ &$\Ocal^{LR}_2(\mu)$ & $\Ocal^{SLL}_1(\mu)$&$\Ocal^{SLL}_2(\mu)$ \\  \hline\hline
$K\text{-}\ol{K}$         & -0.159& 0.261& -0.0761   & -0.132 \\ 
$B_d\text{-}\ol{B}_d$ & -0.186 & 0.241& -0.0909 & -0.167 \\
\hline
\end{tabular}
\end{table}
We shall focus on flavor violating processes in the quark sector.
Flavor violating effects induced by scalar fields are summarized in Ref.~\cite{Buras:2013rqa}. 
The flavons would affect also to the neutral meson mixing. 
For the $K$-$\ol{K}$, $B_d$-$\ol{B}_d$ mixing, the relevant observables are defined as 
\begin{align}
\eps_K  = \frac{\kappa_\eps e^{i\phi_\eps}}{\sqrt{2} \Delta M_K }  \mathrm{Im}\left(M_{12}(K)\right), 
\quad 
\Delta M_{d} = 2 \abs{M_{12}(B_d)}, 
\end{align}
where $\kappa_\eps = 0.94\pm 0.02$, 
$\phi_\eps = (43.51\pm 0.05)^\circ$~\cite{Buras:2008nn,Buras:2010pza}
and $\Delta M_K = 0.005293$ ps$^{-1}$~\cite{Tanabashi:2018oca}.   
The off-diagonal matrix elements are given by 
\begin{align}
M_{12}(K) =&\ M_{12}^\mathrm{SM}(K) 
- \sum_{\scal=\sigma, a} \frac{1}{4m_\scal^2} 
                                   \left[ c_{LL}^K(m_\scal)  
                                               \left\{ \left(\hla^{d,\scal}_{21}\right)^2 
                                          +\left(\hla^{d,\scal *}_{12}\right)^{2} \right\}
                      + 2 c_{LR}^K  
                           \hla^{d,\scal}_{21} \hla^{d,\scal * }_{12}\right],  \\
M_{12}(B_d) =&\ M_{12}^\mathrm{SM}(B_d) 
- \sum_{\scal=\sigma, a} \frac{1}{4m_\scal^2} 
                                   \left[ c_{LL}^{B_d}(m_\scal)   
                                       \left\{ \left(\hla^{d_\scal}_{31}\right)^2 +\left(\hla^{d,\scal *}_{13}\right)^{2} \right\}
                                       + 2  c_{LR}^{B_d} 
                                           \hla^{d,\scal}_{31} \hla^{d,\scal * }_{13}\right],   
\end{align}
where $M_{12}^\mathrm{SM}(M)$ denotes the SM contributions to the meson $M$.
The coefficients of $c^M_{LL}$ and $c^M_{LR}$ ($M=K, B_d$) are given by   
\begin{align}
 c_{LL}^M(m_\scal) =&\ \left[1+\frac{\alpha_s}{4\pi}\left(-3\log{\frac{m_\scal^2}{\mu^2}}+\frac{9}{2}\right)\right]
                              \Ocal^{\mathrm{SLL}}_1(\mu) 
                             + \frac{\alpha_s}{4\pi} \left(-\frac{1}{12}\log\frac{m_\scal^2}{\mu^2}+ \frac{1}{8}\right)  
                             \Ocal^{\mathrm{ SLL}}_2(\mu),   \\
 c_{LR}^M=&\  -\frac{3}{2} \frac{\alpha_s}{4\pi} \Ocal^{\mathrm{LR}}_1(\mu)  
                                +\left(1-\frac{\alpha_s}{4\pi} \right) \Ocal^{\mathrm{LR}}_2(\mu).   
\end{align}
Here, $O_1^{VLL}(\mu)$ and $O_2^{LR}(\mu)$ are the values of hadronic matrices, 
where the renormalization scale $\mu$ is fixed at $1$ TeV in our analysis. 
The QCD corrections accompanied with $\alpha_s$ 
are calculated in Ref.~\cite{Buras:2012fs}. 
We employed the same constant values of the SM contributions and the hadronic matrix elements as in Ref.~\cite{Kawamura:2019rth,Kawamura:2019hxp}. 
Their values relevant to our analysis are shown in Table~\ref{tab-valOs}.

To estimate flavon contributions,
we define ratios of new physics to the SM values as 
\begin{align}
 R_{\eps_K}=
             \frac{\mathrm{Im} \left(M_{12}(K)-M^\mathrm{SM}_{12}(K)\right)}{\mathrm{Im}\left( M^\mathrm{SM}_{12}(K)\right)}, 
\quad 
 R_{B_d}= \abs{\frac{M_{12}(K)-M^\mathrm{SM}_{12}(K)}{ M^\mathrm{SM}_{12}(K)}}. 
\end{align}
In our cases, these are estimated as 
\begin{align}
\label{eq-RepsK}
R_{\eps_K}
=&\ \sum_{\scal=\sigma, a } \frac{10^{15}\ \mathrm{GeV}^2 }{m_\scal^2} 
  \mathrm{Im} \left[1.7 \cdot \left\{\left(\hat{\la}^{d,\scal}_{21} \right)^2 
                                + \left(\hat{\la}^{d,\scal*}_{12}\right)^2 \right\}   
  - 11.  \cdot \hat{\la}^{d,\scal}_{21} \hat{\la}^{d,\scal*}_{12} \right] , \\
\label{eq-RBd}
R_{B_d}
=&\ \sum_{\scal=\sigma, a } \frac{10^{11}\ \mathrm{GeV}^2 }{m_\scal^2} 
\abs{1.1 \cdot \left\{\left(\hat{\la}^{d,\scal}_{31} \right)^2 
                                + \left(\hat{\la}^{d,\scal*}_{13}\right)^2 \right\}   
  - 5.9  \cdot \hat{\la}^{d,\scal}_{31} \hat{\la}^{d,\scal*}_{13} },  
\end{align}
where the QCD corrections are neglected. 
The above parameters are estimated as 
\begin{align}
 R_{\eps_K} \sim&\ 10^{-2} \times 
 \left(\frac{1\ \mathrm{TeV}}{v_s}\right)^2 \left(\frac{100\ \mathrm{GeV}}{m_a} \right)^2 
 \left(\frac{\eps}{0.02}\right)^6, \\ 
 R_{B_d} \sim&\ 10^{-2} \times 
 \left(\frac{1\ \mathrm{TeV}}{v_s}\right)^2 \left(\frac{100\ \mathrm{GeV}}{m_a} \right)^2 
 \left(\frac{\eps}{0.02}\right)^4, 
\end{align}
where the $\order{1}$ coefficients $\rho_d$ are set to be unity. 
The left-left contribution, first term in Eq.~\eqref{eq-RepsK}, 
gives the dominant contribution, 
since $\hat{\lambda}^d_{12} \sim \eps^3$ and $\hat{\lambda}^d_{21} \sim \eps^4$.
All the contributions are sizable for $B_d$-$\ol{B}_d$ mixing 
since $\hat{\lambda}^d_{13} \sim \hat{\lambda}^d_{31} \sim \eps^2$. 
Thus the flavon contributions could affect to the observables 
at a few percent level against the SM values 
for a larger $\hat{\lambda}^{d,\scal}$, a small VEV or a light flavon.

Experiments measure $\eps_K$ so precisely 
that the error is dominated by the theoretical ones,  
such as determination of hadron matrix elements, CKM matrix elements in the SM. 
The error bar is about $10\ \%$. 
The limits from $B_d$-$\ol{B}_d$ mixing give similar bound as $\eps_K$. 
We also checked that a constraint from $D$-$\ol{D}$ mixing is weaker, 
and the $B_s$-$\ol{B}_s$ mixing, leptonic decays of 
$B_s\to \mu\mu$ and $K_L \to \mu\mu$ 
give no significant constraints.

\subsubsection{Collider physics} 
\begin{figure}[t]
\centering 
\begin{minipage}[c]{0.48\hsize}
 \centering
\includegraphics[height=70mm]{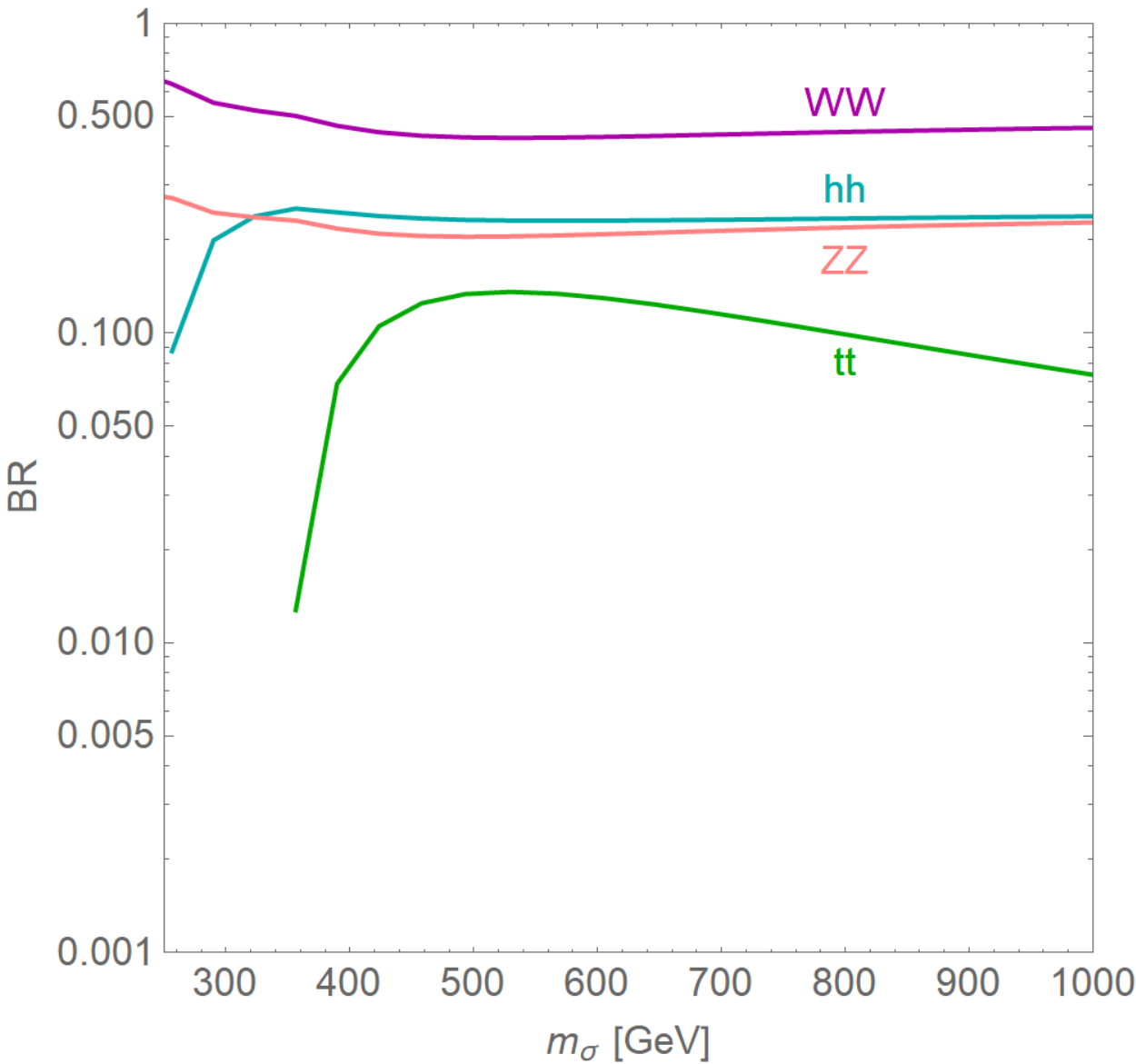}
\end{minipage}
\begin{minipage}[c]{0.48\hsize}
 \centering
\includegraphics[height=70mm]{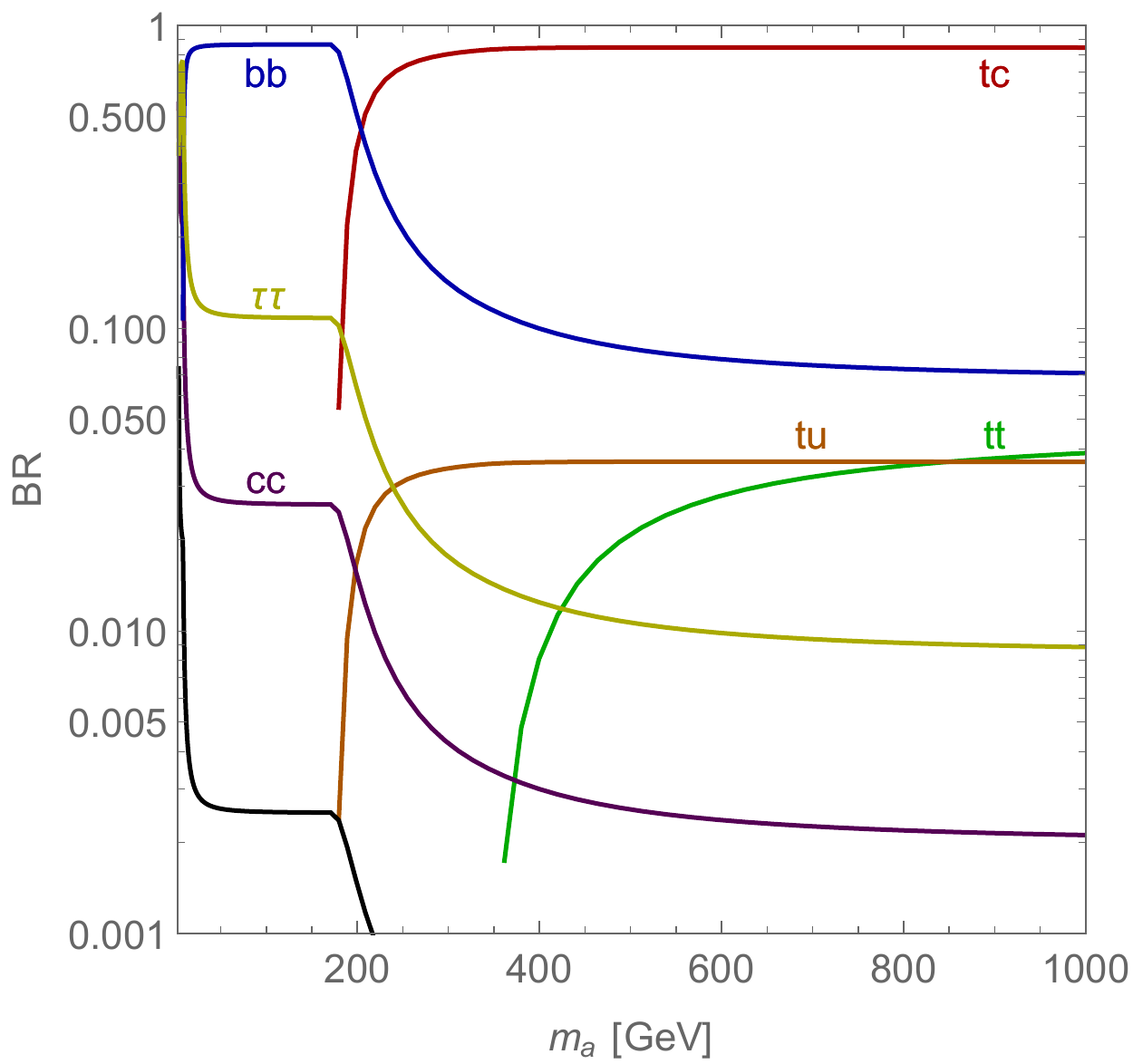}
\end{minipage}
 \caption{
\label{fig-flavondec} 
The branching fractions of the CP-even (left) and CP-odd (right) flavon. 
} 
\end{figure}
We shall discuss collider physics associated with flavons.
Figure~\ref{fig-flavondec} shows branching fractions of the CP-even flavon (left panel) 
and CP-odd flavon (right panel). 
The parameters are fixed at $\tan\beta = 5$, $\eps v_s = 2.0$ TeV, $A_H = 3.0$ TeV and $c_N = c_m =1$.
$A_S$ is scanned to change the flavon masses.
We used the benchmark values of the $\order{1}$ coefficients for the Yukawa couplings shown in Appendix~\ref{App-Bench}.
In addition to the flavon decays to a pair of fermions and vector bosons, 
tree-level decays to bosons of $\sigma \to h h$, $\sigma \to a  Z$, 
$\sigma \to a a $, $a \to \sigma Z$ 
and loop-induced decays of $\sigma/a \to\gamma\gamma, gg$ are taken into account.  
The black line in the right panel is the sum of the remaining branching fractions
not shown in the figure. 
These branching fractions are sub-dominant. 
The decays of flavons, $\sigma \to WW, ZZ, hh$ and $\sigma/a \to t\ol{t}$,
are induced by mixing with the Higgs doublets. 
The dominant decay modes of the CP-even flavon are induced by the couplings not suppressed by $\eps$.  
Since the mixing of the CP-odd flavon to the Higgs doublets are more suppressed by $1/\tan^2\beta$, 
the CP-odd flavon dominantly decays to a pair of fermions through the Yukawa couplings $\hat{\lambda}^{f,a}$.  
In this sense, the CP-even flavon is similar to the Higgs boson 
due to a large mixing, while the CP-odd flavon seems to be a conventional flavon. 
The CP-odd flavon dominantly decays to $tc$ for $m_a \gtrsim m_t$,
while it decays to $bb$ and $\tau\tau$ for $m_a \lesssim m_t$. 
It is noted that $\Gamma(a \to tc)/\Gamma(a \to tt) \sim \epsilon \tan^2 \beta $,
hence $a \to tt$ can be the main decay mode for a smaller $\tan \beta$.
The CP-even flavon dominantly decays to a pair of EW gauge bosons 
as far as these are kinematically allowed. 
For $m_\sigma \lesssim 2 m_W$,  
there is a substantial flavon mixing to the SM Higgs boson 
and the decay modes of $\sigma$ will be similar to that of the SM Higgs boson.  
We could find signals of the mixing with the SM Higgs boson, 
but this happens only if $\eps v_s \sim m_h$ 
or the first two terms in Eq.~\eqref{eq-mhs} are canceled out. 
Constraints on a sizable flavon mixing to the SM Higgs boson 
are studied in Ref.~\cite{Berger:2014gga}. 
The LHC searches with $100$ fb$^{-1}$ data and $\sqrt{s}=14$ TeV 
will constrain parameter space of the mixing
for $\left[ R_S\right]_{31} \gtrsim \order{\sqrt{0.1}}$. 
In this model, $\left[ R_S\right]_{31} \sim v_H/v_s \lesssim 0.04$ for $v_s \gtrsim 5.0$ TeV, 
hence the mixing angle is too small to be detected. 
Thus the mixing between the CP-even scalars are hardly probed at the LHC, 
even if the CP-even flavon is as light as the SM Higgs boson.

As pointed out in Ref.~\cite{Tsumura:2009yf}, 
a flavor violating decay of top quark, $t \to \scal c$ ($\scal= \sigma, a$) 
will be detectable at collider experiments for $m_\scal \lesssim m_t$.
Such a branching fraction is given by 
\begin{align}
 \mathrm{Br}\left(t \to \scal c \right) 
=&\  \frac{m_t}{64\pi \Gamma_t} 
 \sum_{\scal=\sigma, a }
 \left(\abs{\hat{\la}^{u,\scal}_{23}}^2+\abs{\hat{\la}^{u,\scal}_{32}}^2 \right)  
 \left(1-\frac{m_\scal^2}{m_t^2}\right)^2 \notag \\ 
\sim&\  
7\times 10^{-8}\times \left(\frac{10\ \mathrm{TeV}}{v_s} \right)^2\left(\frac{\eps}{0.02}\right)^2,    
\end{align}
where the charm mass is neglected. $\Gamma_t$ is the decay width of the top quark. 
In the second equality, the decay $t\to \sigma c$ and the CP-odd flavon mass are neglected. 
The future sensitivity at 100 TeV hadron collider is 
$2.2 \times 10^{-6}$~\cite{AguilarSaavedra:2000aj,Bauer:2016rxs}.
The flavor changing coupling also predicts same-sign top signal, 
$pp\to ta \to tt \ol{c}$, 
but this is not accessible when $v_s \gtrsim 2.0$ TeV~\cite{Bauer:2016rxs} 
in order to realize $\mueff \gtrsim 90$ GeV.

\subsection{Numerical result}  
\begin{figure}[t]
\centering
\includegraphics[height=100mm]{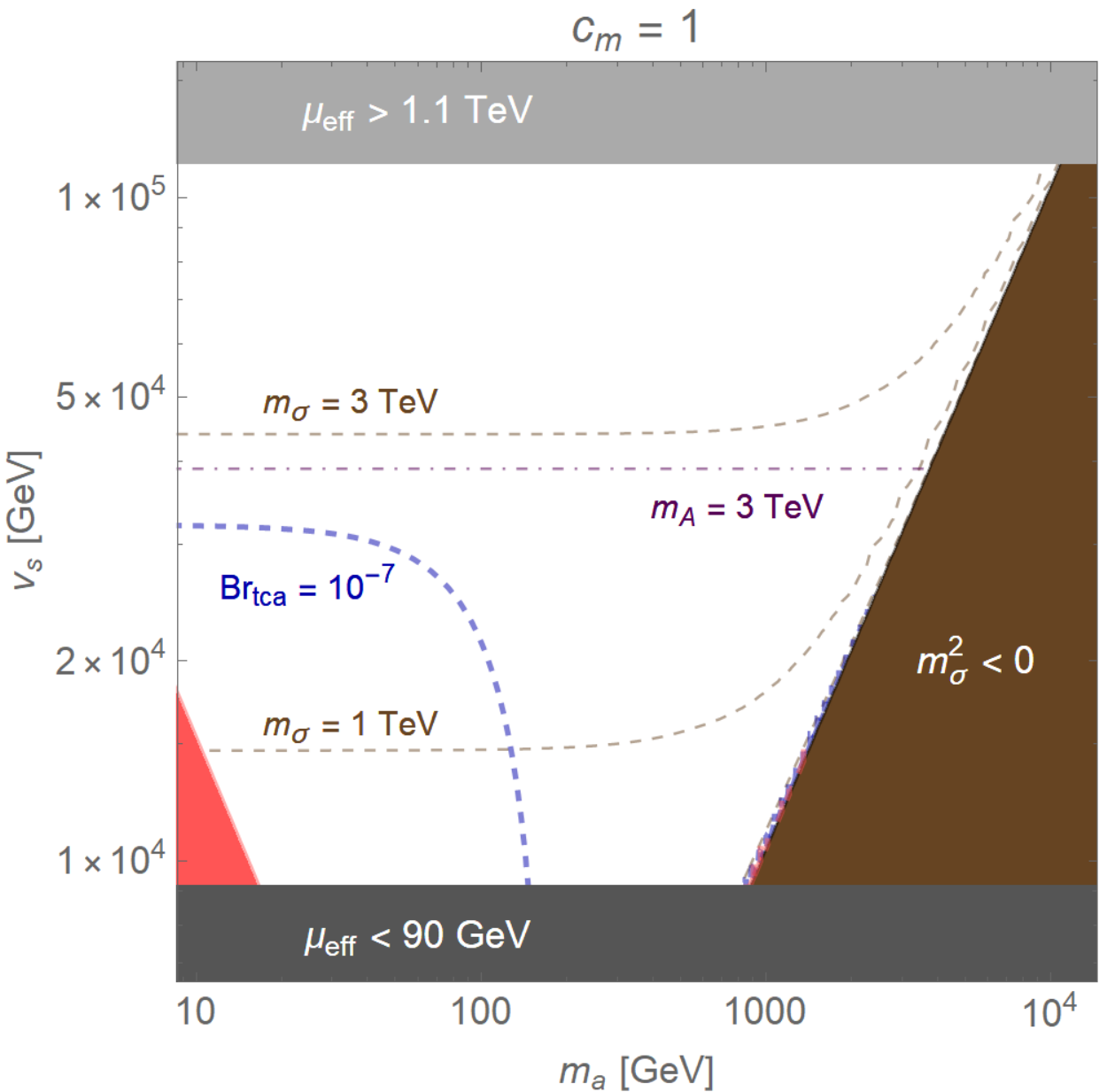}
\caption{\label{fig-rslt1}  
Allowed region in the ($m_a$, $v_s$)  [GeV] plane for $c_m = 1$ 
in the Higgsino LSP case.  
In the wino LSP case, a larger VEV $v_s$ is allowed.
White region is consistent with observations. 
$\abs{R_{\eps_K}} > 0.1$ in the red region. 
The blue lines show the branching fraction of the flavor changing top decay $t\to ca$ is at $10^{-7}$. 
The other dashed lines show masses of the CP-even flavon and CP-odd Higgs boson.  
}
\end{figure}

\begin{figure}[t]
\centering
\includegraphics[height=100mm]{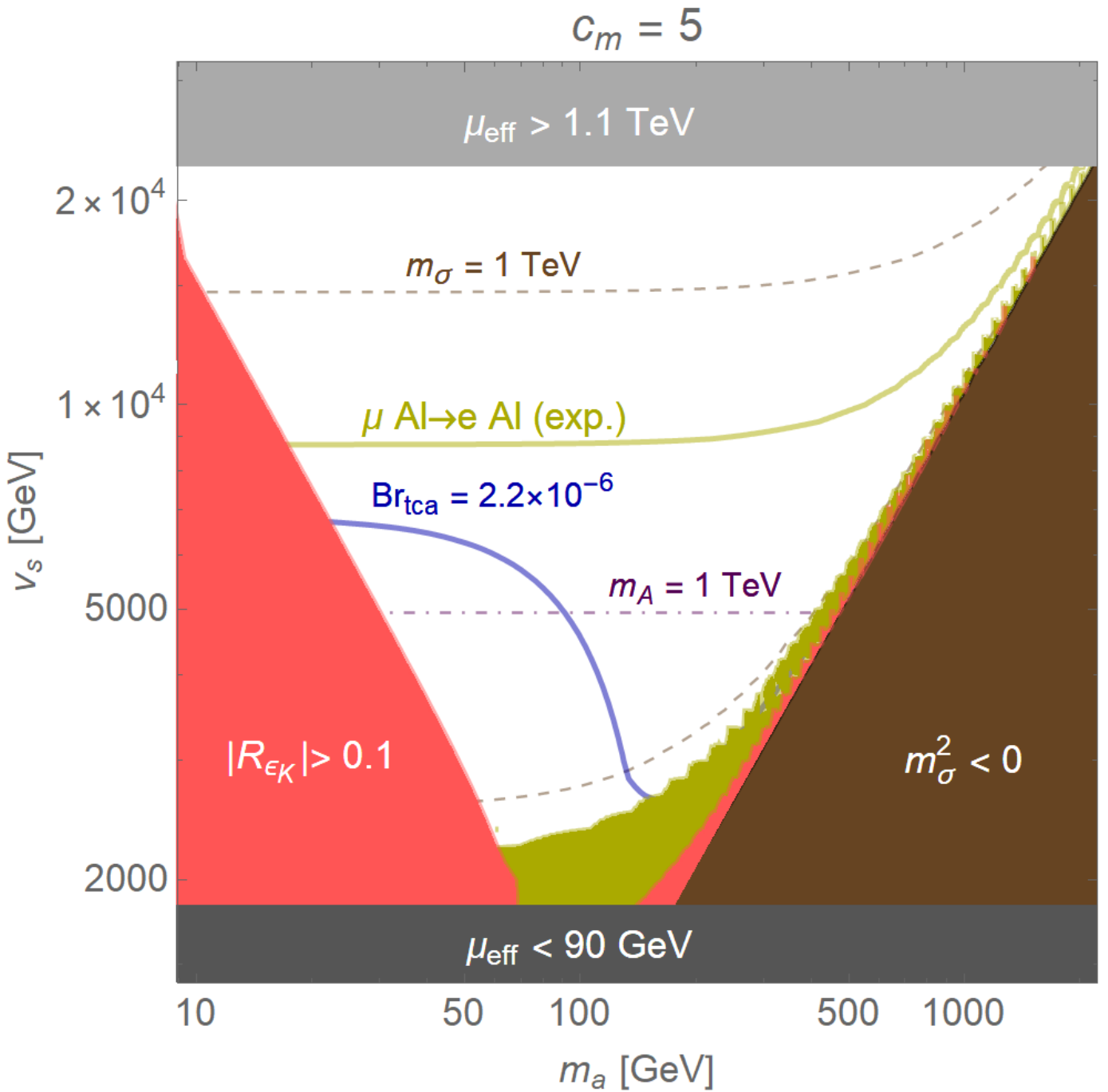}
\caption{
\label{fig-rslt5}
Similar figure to Fig.~\ref{fig-rslt1} but for $c_m = 5$.
White region is consistent with observations. 
The yellow region is excluded by $\mu\to e$ conversion.  
The brown dashed line near the bottom indicates $m_\sigma = m_t$. 
}
\end{figure}

Figure~\ref{fig-rslt1} shows the allowed parameter space on ($m_a$, $v_s$) plane
for $\tan\beta = 5$, $A_H$ = 3.0 TeV and $c_N = c_m =1$
in the Higgsino LSP case. 
The values of Yukawa couplings shown in Appendix~\ref{App-Bench} are used.  
The white region is allowed by current experiments. 
In the dark gray region, the Higgsino is lighter than the experiments bound $\sim 90$ GeV.
In the light gray region, 
the Higgsino is too heavy and its relic density will over-close the universe.
In the wino LSP case, larger VEV $v_s$ is allowed.
The CP-even flavon is tachyonic in brown region. 
In the red region,  $\abs{R_{\eps_K}} > 0.1$ 
and the flavon contributes to $\eps_K$ more than $10$\% 
against the SM contribution.   
It is noted that
there exists the red region also near the region of $m_\sigma^2 < 0$, 
where the CP-even flavon is very light.  Such a region 
is very narrow to be seen. 
A light CP-even flavon with $m_\sigma \ll \order{\eps v_s}$ is owing to
a cancellation between the two terms in Eq.~\eqref{eq-mhs}. 
The dashed lines show the CP-even flavon masses 
and the dot dashed line indicates the CP-odd Higgs mass. 
The blue dashed line shows $\mathrm{Br}\left(t\to ac\right) = 10^{-7}$. 
There is no parameter space where $\mathrm{Br}\left(t\to ac\right)$  
is larger than the future sensitivity at the 100 TeV collider.  
Thus, vast parameter space will not be constrained 
by measurements of the processes induced by the flavons.

Figure~\ref{fig-rslt5} is the same as Fig.~\ref{fig-rslt1}, but $c_m = 5$
in the Higgsino LSP case. 
The flavon VEV shown in this figure is lower than cases with $c_m = 1$, 
since the Higgsino mass of $\mu_{\rm eff} \sim c_m/2 \cdot \epsilon v_s$
linearly depends on $c_m$.
Thus the Yukawa couplings of the flavons $\sim v_H/v_s $
become larger than those in Fig.~\ref{fig-rslt1}.
The red and yellow regions are excluded by the current limits from $\eps_K$ and $\mu\to e$ conversion, 
respectively. 
The dashed line in the bottom indicates $m_\sigma = m_t$,  
and the top quark can decay into the CP-even flavon below this line. 
The region below the blue line will be covered by the future 100 TeV collider. 
Furthermore, future measurements for $\mu\to e$ conversion 
will probe the region lower than the yellow line. 
Thus the wide parameter space will be tested by the future experiments in this case.

\section{Summary and discussion} 
\label{sec-concl} 
In this paper, we proposed a model with 
the Froggatt-Nielsen mechanism controlled by a $\Zn{N}^F$ symmetry,
in which a flavon field explains not only flavor hierarchy but also
the size of Higgsino mass as a solution for the $\mu$-problem in the MSSM.   
The Higgsino is a well motivated candidate for the DM, 
since the thermal relic is explained consistently with the null result in direct detections. 
Furthermore, the abelian flavor $\Zn{N}^F$ symmetry for the FN mechanism 
also regulates a structure of the Higgs and flavon potential. 
Thus the origin of fermion mass hierarchy is closely related 
to DM physics and Higgs physics.

We found charge assignments of the discrete flavor symmetry, 
which explains the fermion mass hierarchy and   
does not have anomalies in the non-abelian gauge groups of the SM in a case of $N=4$. 
Together with the condition to prevent the existence of 
an exotic minimum deeper than the EW vacuum, 
the power of $m=2$ in the superpotential $W \ni (S^m/\Lambda^{m-1}) H_u H_d$
is uniquely determined. 
As a consequence, $\tan\beta \sim \order{1}$ is required to explain 
the observed bottom to top quark mass ratio through the anomaly conditions. 
Our analysis for the Higgs potential shows 
that the realistic EW vacuum can be realized
even if the flavon direction has only cutoff suppressed couplings. 
It is also interesting that CP-odd Higgs boson mass
have the upper bounds, so that the EW vacuum is the deepest minimum.

The large portion of parameter space is allowed by the experiments. 
In the Higgsino LSP case,
the flavon VEV is constrained from above to explain the Higgsino relic density,
and also restricted from below to be consistent with the collider bound on the chargino. 
In the wino LSP case, the upper bound on $v_s$ is relaxed.
The flavon VEV is related also to
the Higgs boson and flavon masses. 
Altogether, there is a window in the parameter space 
consistent with DM particle and EW vacuum. 
The flavor constraints are not severe
because the flavor violating couplings are suppressed 
by a large flavon VEV, $v_s \gtrsim 10$ TeV.   
Only the restricted parameter space where there exist 
light flavons or large Yukawa couplings 
will be covered by future experiments.

We shall give comments about larger discrete flavor symmetry, namely $N>4$.
According to Eq.~\eqref{eq-epsN}, the popular choice of $\eps\sim 0.22$ is realized when $N = 9$.
Of course, choices of $\order{1}$ coefficients can change the relation between the value of $\eps$
and top to up quark mass ratio.
Hence, $\eps \approx 0.22$ can be obtained for the realistic mass hierarchies when 
$N \ge 6$,
while a smaller  $\eps = {\cal O}(0.1)$ may be found for $N=5$. 
For larger $N$, there would be more ambiguities of textures in the Yukawa matrices 
and the discussions about the charge assignments would not be as rigid as 
the case of $N=4$ studied in this paper.
We may find also a variety of choices of charge assignments and $\order{1}$ coefficients consistent with the realistic fermion hierarchies as well as 
anomaly cancellation conditions for larger $N$.
In these cases, the powers of Higgs to flavon coupling, $m$,
may not be uniquely fixed because of the ambiguities.
A relation between $N$ and $m$ will significantly change the Higgs and DM physics.
For example, the Higgsino-like DM is expected for $m-1 > N-3$,
while the singlino (flavino)-like DM is expected for $m-1 < N-3$. 
For $m-1=N-3$, whether the Higgsino or the singlino become DM 
depends on a choice of ${\cal O}(1)$ coefficients.
The qualitative features in flavor physics will be similar for larger $N$ cases,
and flavor violating processes are strongly suppressed by the Yukawa couplings.
A specific feature only in $N=4$ would be the alignment of the Higgs and flavon Yukawa matrices
owing to the non-hierarchical texture of $Y_e$ for muon and tau as shown 
in Eq.~\eqref{eq-nYsamp}.
This strongly suppresses the LFV couplings of the flavon as in Eq.~\eqref{eq-flavonY}.
The LFV processes would be more relevant for larger $N$.

In this paper, we do not discuss flavor violations induced by SUSY breaking. 
Since we have mentioned about the flavor structure of the fermions, 
we may be able to address those in soft SUSY breaking. 
A choice of discrete charge determines also hierarchy in the soft mass.
For instance, 
the soft masses of the right-handed sfermions between first and second generation,
$\tilde{u}^\dag_1 \tilde{u}_2,~ \tilde{d}^\dag_1 \tilde{d}_2,~ \tilde{e}^\dag_1 \tilde{e}_2$,
would be suppressed by $\eps$ while those of the left-handed sfermions,
$\tilde{Q}^\dag_1 \tilde{Q}_2,~ \tilde{L}^\dag_1 \tilde{L}_2$,
would not be in our model.
Flavor violation from SUSY breaking as well as 
higher dimensional operators from the \Kahler\ potential
may open new possibilities to probe this model as discussed in Appendix~\ref{KYukawa}. 
This is left as future work.

\section*{Acknowledgment} 
This work is supported by the Ministry of Education,
Culture, Sports, Science (MEXT)-Supported Program
for the Strategic Research Foundation at Private Universities 
``Topological Science'' Grant No.\ S1511006.
The work of J.K.\ is supported in part by the Department of Energy (DOE) under Award No.\ DE-SC0011726 and the Grant-in-Aid for Scientific Research from the
Ministry of Education, Science, Sports and Culture (MEXT), Japan No.\ 18K13534.

\appendix
\section{Analytical formulas} 
\label{App-anal}
The minimization condition for a EW symmetry breaking minimum is given by 
\begin{align}
&m_{H_d^2} + \abs{\mueff}^2 + g^2 v_H^2 c_{2\beta} -\mueff\Beff \tan\beta + (c_m\eps^{m-1}v_u)^2 
= 0, \\
&m_{H_u^2} + \abs{\mueff}^2 - g^2 v_H^2 c_{2\beta} -\mueff\Beff \cot\beta + (c_m\eps^{m-1}v_d)^2  
= 0, \\
& 
m_S^2 + (N-1) \left(c_N \eps^{N-3} v_s \right)^2 + A_S \eps^{N-3} v_s  
                          - A_H  \eps^{m-1} \frac{v_uv_d}{v_s}  \\ \notag 
&\quad     + \frac{1}{m} \left(\laeff v_H \right)^2  -(N+m-2) c_N c_m \eps^{N+m-4} v_u v_d 
      + (m-1) \left(c_m \eps^{m-1}\frac{v_u v_d}{v_s} \right)^2 = 0, 
\end{align}
where $v_H^2 := v_u^2+v_d^2$, 
$\tan\beta := v_u/v_d$,  $\laeff := c_m \eps^{m-1}$, 
$\mu_{\mathrm{eff}} := c_m/m \cdot \eps^{m-1} v_s$ and  
$B_\mathrm{eff} := A_H /c_m + c_N m\eps^{N-3} v_s$.

The CP-even Higgs mass matrix is given by 
\begin{align}
 \Mcal^2_{S,11} =&\   2g^2 v_d^2 + \mueff \Beff \tan\beta,  \\
 \Mcal^2_{S,12} =&\   2(\laeff^2 -g^2) v_uv_d - \mueff\Beff,  \\
 \Mcal^2_{S,22} =&\ 2g^2 v_u^2 + \mueff \Beff \cot\beta,  \\
 \Mcal^2_{S,13} =&\  -\eps^{m-1} v_u \left[A_H+(N+m-2)c_Nc_m\eps^{N-3}v_s \right]   
                              +2\laeff  \mueff v_d  +\order{v_H^3/v_s}, \\
 \Mcal^2_{S,23} =&\ -\eps^{m-1} v_d \left[A_H+(N+m-2)c_Nc_m\eps^{N-3}v_s \right]   
                              +2 \laeff \mueff v_u + \order{v_H^3/v_s},  \\
 \Mcal^2_{S,33} =&\  (N-2)\eps^{N-3} A_S v_s + 2(N-1)(N-2)(c_N \eps^{N-3}v_s)^2 + \order{v_H^2},  
\end{align}
The CP-odd Higgs mass matrix is given by 
\begin{align}
 \Mcal^2_{P,11} =&\  \mueff\Beff\tan\beta,\quad  \Mcal^2_{P,22} = \mueff\Beff\cot\beta, \quad    
 \Mcal^2_{P,12} = \mueff \Beff, \\ 
 \Mcal^2_{P,13} =&\  m \mueff\left(\Beff - N c_N \eps^{N-3}v_s \right)\cdot\frac{v_u}{v_s},\quad  
 \Mcal^2_{P,23} =   m \mueff\left(\Beff - N c_N \eps^{N-3}v_s \right)\cdot\frac{v_d}{v_s}, \\
 \Mcal^2_{P,33} =&\  -A_S N  \eps^{N-3}v_s + 
                               (N-m)^2 c_N c_m \eps^{N+m-4} v_u v_d 
                             +A_H m\eps^{m-1}  \frac{v_uv_d}{v_s}.  
\end{align}
The CP-odd doublet Higgs mass squared is positive if $\mueff \Beff >0$.  
These matrices are approximately diagonalized by 
\begin{align}
 R_S^0 = 
\begin{pmatrix}
 c_\beta & -s_\beta &0 \\ 
s_\beta & c_\beta & 0 \\
0 & 0 & 1 \\
\end{pmatrix},
\quad 
 R_P^0 = 
\begin{pmatrix}
 c_\beta & s_\beta &0 \\ 
-s_\beta & c_\beta & 0 \\
0 & 0 & 1 \\
\end{pmatrix}.  
\end{align}
After the rotation, the CP-even matrix 
$\tilde{\Mcal}_S^2 := R_S^{0T}\Mcal_S^2R_S^0 $ becomes 
\begin{align}
\tMcal_{S,11}^2 
=&\ \left(2 g^2 \cos^2 2\beta  +\laeff^2 \sin^2 2\beta \right) v_H^2, \\ 
\tMcal_{S,12}^2 
=&\ \frac{1}{2}\left(\laeff^2-2g^2 \right) v_H^2 \sin 4\beta, \\ 
\tMcal_{S,22}^2 
=&\ 2 \mueff \Beff / \sin2\beta, \\  
\tMcal_{S,13}^2  
=&\ -\eps^{m-1} v_H \left(A_H +(N+m-2) c_Nc_m \eps^{N-3}v_s\right) \sin{2\beta}  + 2\laeff \mueff v_H, \\  
\tMcal_{S,23}^2 
=&\ -\eps^{m-1}v_H \left(A_H +(N+m-2) c_Nc_m \eps^{N-3}v_s\right) \cos{2\beta}, 
\end{align}
and $\tMcal_{S,33} = \Mcal_{S,33}$. 
When $\tMcal_{S,12}^2, \tMcal_{S,23}^2 \ll \tMcal_{S,22}^2$,  
and $\tMcal_{S,13}^2, \ll \tMcal_{S,33}^2 $,  
the rotation matrix is approximately given by 
\begin{align}
 R_S \sim 
\begin{pmatrix}
\cos \beta - \delta_3 \sin\beta & -\sin\beta- \delta_3 \cos\beta  &\delta_1 \sin\beta -\delta_2 \cos\beta \\
\sin\beta+\delta_3 \cos\beta & \cos\beta-\delta_3 \sin\beta & -\delta_1 \cos\beta -\delta_2 \sin\beta \\ 
\delta_2 & \delta_1 & 1 
\end{pmatrix}
+ \order{\delta_i^2}, 
\end{align}
where 
\begin{align}
 \delta_1 := 
\frac{ \tMcal_{S,23}^2}{  \tMcal_{S,22}^2 -\tMcal_{S,33}^2 },\quad 
 \delta_2 := 
\frac{\tMcal_{S,13}^2 }{ \tMcal_{S,11}^2 - \tMcal_{S,33}^2 },\quad 
 \delta_3 := 
\frac{ \tMcal_{S,12}^2}{ \tMcal_{S,11}^2 - \tMcal_{S,22}^2  }. 
\end{align}

Similarly, the mixing matrix for the CP-odd mass matrix is given by 
\begin{align}
R_P =&\  
\begin{pmatrix}
 \cos\beta & \sin\beta & \eta \sin\beta   \\ 
 -\sin\beta  & \cos\beta  & \eta \cos\beta  \\
 0 & - \eta &  1 
\end{pmatrix}
+ \order{\eta^2}, \quad 
 \eta = \frac{\Mcal^2_{P,13}\sin\beta + \Mcal^2_{P,23} \cos\beta}{ \Mcal^2_{P,33}- 2\mueff \Beff /\sin{2\beta} }. 
\end{align}
In the case of $N=4$, $m=2$, 
\begin{align}
\delta_1 \sim \eta \sim \order{\frac{v_H}{v_s }\cot\beta},\quad 
\delta_2 \sim \order{\frac{v_H}{v_s }}.
\end{align}

The widths of the flavon decays are given by  
\begin{align}
 \Gamma \left(a_i \to h_j Z \right) 
=&\ \frac{m_{a_i}^3}{32\pi v_H^2} 
   \abs{ \left[R_S\right]_{1j}\left[R_P\right]_{1i}
            -\left[R_S\right]_{2j}\left[R_P\right]_{2i} }^2  \\ \notag 
&\ \quad \times   \left(1- 2 \frac{m_{h_j}^2 + m_Z^2}{m_{a_i}^2} 
     + \frac{(m_{h_j}^2-m_Z^2)^2}{m_{a_i}^4} \right)^{3/2}, \\
 \Gamma \left(h_i \to a_j Z \right) 
=&\ \frac{m_{h_i}^3}{32\pi v_H^2} 
   \abs{ \left[R_S\right]_{1i}\left[R_P\right]_{1j}
            -\left[R_S\right]_{2i}\left[R_P\right]_{2j} }^2  \\ \notag 
&\ \quad\times   \left(1- 2 \frac{m_{a_j}^2 + m_Z^2}{m_{h_i}^2} 
     + \frac{(m_{a_j}^2-m_Z^2)^2}{m_{h_i}^4} \right)^{3/2},   \\
 \Gamma \left(h_i \to  V V \right) 
=&\ \frac{ \kappa_V m_{h_i}^3}{32\pi v_H^2} 
   \abs{ c_\beta \left[R_S\right]_{1i} + s_\beta\left[R_S\right]_{2i} }^2 
   \sqrt{1-4\frac{ m_V^2}{m_{h_i^2}}} 
  \left(1-4\frac{ m_V^2}{m_{h_i^2}}+12 \frac{m_V^4}{m_{h_i^4}}  \right), \\
 \Gamma \left( \scal \to \phi \phi    \right) 
=&\ \frac{ \abs{A_{\scal \phi \phi}}^2  }{32\pi m_{\scal}}    \sqrt{1-4\frac{ m_\phi^2}{m_{\scal}^2 }},  \\
 \Gamma \left( \scal\to f_i \ol{f}_j  \right) 
=&\ N_c^{f} \frac{m_\scal}{32\pi} \sqrt{1-2\frac{m_{f_i}^2+m_{f_j}^2}{m_\scal^2}+4\frac{(m_{f_i}^2-m_{f_j}^2)^2}{m_\scal^4}} \\ 
 \notag &\quad \times 
         \left[\left(\abs{\hla^{f,\scal}_{ij}}^2+\abs{\hla^{f,\scal}_{ji}}^2\right)\left(1-\frac{m_{f_i}^2+m_{f_j}^2}{m_\scal^2} \right) 
                  -\mathrm{Re} \left(\hla^{f,\scal}_{ij}\hla^{f,\scal*}_{ji} \right)  \frac{4 m_{f_i} m_{f_j}}{m_\scal^2} \right], 
\end{align}
where $\kappa_Z = 1/2$ and $\kappa_W = 1$ for $V= Z,W$.  
Here, $\scal, \phi = h_i, a_i$ and $N_c^f = 3~(1)$ for quarks~(leptons). 
The trilinear coupling can be obtained by 
\begin{align}
 A_{h_i h_j h_k} = \left.\frac{\partial^3 V}{\partial h_i \partial h_j \partial h_k}\right|_\mathrm{min},\quad  
 A_{h_i a_j a_k} = \left.\frac{\partial^3 V}{\partial h_i \partial a_j \partial a_k}\right|_\mathrm{min},   
\end{align}
where $h_i = (h,H,\sigma)$ and $a_i = (A, a)$. 
Here, $|_\mathrm{min}$ means that the fields should be replaced by their VEVs after differentiations. 
When the mixing between the Higgs bosons only via
$R_S^0$ and $R_P^0$ are taken into account,  
the relevant trilinear couplings are given by 
\begin{align}
 A_{\sigma hh} \sim&\ \frac{1}{\sqrt{2}} 
                             \left[ c_m^2 \eps^2 v_s 
                                        -\left(\eps A_H+4c_N c_m \eps^2 v_s\right) \sin{2\beta} 
                         +3c_m^2 \eta_H^2v_s \sin^2{2\beta} \right] 
     \sim\order{\eps^2 v_s}, \\ 
 A_{h \sigma \sigma} \sim&\ \frac{v_H}{\sqrt{2}}  \left[ 
3c_m^2 \eps^2 
 - \left(\dfrac{A_H}{\Lambda}+ 12 c_N c_m \eps^2\right) \sin{2\beta}
+  c_m^2 \eta_H^2 \sin^2{2\beta}
  \right] \sim \order{\eps^2 v_H } , \\
 A_{haa} \sim&\ \frac{v_H}{\sqrt{2}} 
                             \left[ c_m^2 \eps^2  + \frac{A_H}{\Lambda} \sin{2\beta}
                                       +  c_m^2 \eta_H^2 \sin^2{2\beta}  \right] \sim \order{\eps^2 v_H} ,   \\
 A_{\sigma aa} \sim&\ \frac{v_s}{\sqrt{2}} 
                             \left(12 c_N^2 \eps^2 -6 \frac{A_S}{\Lambda} 
                                  + c_m^2 \eta_H^2  \right) \sim \order{\eps^2 v_s},   
 \end{align}
where $\eta_H := v_H/\Lambda$. 
The formulas for the loop-induced decays can be found in e.g. Ref.~\cite{Djouadi:2005gi}.

\section{Higher dimensional operators in K$\mathrm{\mathbf{\ddot{a}}}$hler potential} 
\label{KYukawa}
We discuss whether our model is modified by
possible corrections from higher dimensional operators 
in the \Kahler\ potential 
for $N=4$ with $W \ni S^4/\Lambda - (S^m/\Lambda^{m-1}) H_u H_d$. 
Throughout this paper, 
sfermions and gauginos are assumed to be heavier than $10$ TeV
and irrelevant to phenomenology. 
Our conclusion for $m=1$ is not changed by the higher dimensional operators, 
since the problem is that the minimum along the $H_u$ direction, 
$V_{H_u, \mathrm{min}} \sim - \order{v_s^4}$,  
is much deeper than the EW vacuum whose the depth is $\sim -\order{\eps^2 v_s^4}$. 
This can not be changed by higher dimensional operators.

For $m=2$, 
the higher dimensional operators could change the results discussed so far, 
if they affect the hierarchical structure in the Yukawa matrices and/or the Higgs potential. 
There exists $\order{S^3/\Lambda^{3}}$ terms in the Yukawa matrix and
also $\order{|S|^2/\Lambda^{2}}$ terms in the Higgs potential.
Hence, it is sufficient to check 
$\order{\Lambda^{-2}}$ and $\order{\Lambda^{-1}}$ corrections 
associated with the flavon in the \Kahler\ potential  
in order to see 
whether the hierarchical structures in the Yukawa matrices and Higgs potential
are altered by them.

We focus on the Higgs potential first. Because of charge assignment, 
it is impossible to write $\order{\Lambda^{-1}}$ terms in the \Kahler\ potential
made only of $S, H_u$ and $H_d$.
For $\order{\Lambda^{-2}}$ terms,
$K \ni |S|^2 |H_{u,d}|^2/\Lambda^2 + |S|^4/\Lambda^2$ 
change kinetic terms only by $\epsilon^2$. 
These do not alter the hierarchical structure in the Higgs-flavon sector.
For terms associated with SUSY breaking, 
we may have $K \ni (S^\dag)^2 H_u H_d/\Lambda^2$. 
This contributes to the Higgsino mass as 
$\sim \epsilon \langle F_{S}^{\dag}\rangle/\Lambda \sim \eps^3 v_s$, 
where $F_S^\dag \sim S^3/\Lambda$.
This size is negligible to that from the superpotential, $\mu_{\rm eff} \sim \epsilon v_s$.

The terms involving the SM fermions in the \Kahler\ potential are given by 
\begin{align}
\Delta_Q K    =&\ 
\left(
\frac{a^i_j}{\Lambda} S Q_i^\dag Q_j 
+\frac{\tilde{a}^i_j}{\Lambda^2} D^\alpha D_\alpha S \cdot Q_i^\dag  Q_j    
+ \frac{{b}^i_j}{\Lambda^2} S^2 Q_i^\dag Q_j 
+ \frac{c^{ij}}{\Lambda^2} S^\dag H_a Q_i Q_j  
+ h.c. \right)   \notag \\
& \quad + \frac{d^i_j}{\Lambda^2} S^\dag S Q_i^\dag Q_j 
+ \frac{e_{ijkl}}{\Lambda^2} Q_i^\dag Q_j Q_k^\dag   Q_l 
+ \order{\Lambda^{-3}},    
\end{align}
where $Q_i$'s are the quark and lepton chiral multiples 
and $H_a$'s are the Higgs doublets $H_u$ or $H_d$. 
The coupling constants 
$a^i_{j}, \tilde{a}^i_{j}, b^i_{j}, c^{ij}, d^i_{j}$ and $e_{ijkl}$ are $\order{1}$ coefficients. 
Some of them are more suppressed by $\eps = \vev{S}/\Lambda$ 
to make the operators invariant under the $\Zn{4}^F$ symmetry.   
The chiral covariant derivative is defined as 
$D_\alpha := \partial/\partial\theta^\alpha - i\left(\sigma^\mu \theta^\dagger  \right)_\alpha \partial_\mu$. 
Here, the gauge supermultiplets are omitted. 
The gauge interactions of the SM fermions 
will be obtained by replacing a space time derivative $\partial_\mu$ to a gauge covariant one.

The terms proportional to $\tilde{a}^i_j$ and $c^{ij}$ 
may contribute to both kinetic terms and Yukawa couplings. 
The kinetic term corrections also contribute to Yukawa couplings
by canonical normalization (see below).
As shown below, however, the size of corrections turn out to be
$\langle F_S \rangle/\Lambda^2 \sim \eps^3$, which is comparable
to the smallest Yukawa coupling. 
For kinetic term correction, we have 
$D^\alpha D_\alpha S/\Lambda^2 \sim F_S/\Lambda^2$.
For Yukawa coupling correction, it is noted in Section~\ref{sec-fmm} that 
there exists $W \ni (S/\Lambda)^{N-1}H_a Q_i Q_j = (S/\Lambda)^{3}H_a Q_i Q_j $ 
so long as we have $K \ni c^{ij} S^\dag H_a Q_i Q_j/\Lambda^2$. 
Since $\langle F_S^\dag \rangle/\Lambda^2 \sim \eps^3$, 
corrections from the \Kahler\ potential give the same order contribution as that from the superpotential for $N=4$.      
With a general $N$, 
$\langle F_S \rangle/\Lambda^2 \sim \epsilon^{N-1}$ with
$W \ni S^N/\Lambda^{N-3}$ will be similarly satisfied,
where $\epsilon^{N-1}$ is comparable to the smallest Yukawa coupling.

With component fields, the higher dimensional terms are rewritten as
\begin{align}
\label{eq-highq}
\int &\  d^2\theta d^2\theta^\dag   \Delta_Q K =  
\frac{e_{ijkl}}{\Lambda^2} q_i^\dag \ol{\sigma}^\mu q_j \cdot q_k^\dag \ol{\sigma}_\mu  q_l 
\\  \notag 
&\  +\frac{i}{2} \left[\left\{ \frac{a^i_j}{\Lambda} \partial^\mu S
   + 2 \frac{b^i_j}{\Lambda^2} S\partial^\mu S \right.\right. 
 \left.\left. + \frac{d^i_j}{\Lambda^2} 
 \left(S^*\partial^\mu S-S\partial^\mu S^* + i\tilde{S}^\dag \ol{\sigma}^\mu \tilde{S} \right)   \right\}
q_i^\dag \ol{\sigma}_\mu q_j
  \right. \\ \notag 
 &\ \left.  + \left(
\frac{a^i_j}{\Lambda} S+ \frac{b^i_j}{\Lambda^2} S^2  + \frac{d^i_j}{\Lambda^2}\abs{S}^2
\right) 
\left( q^\dag_i \ol{\sigma}^\mu \partial_\mu q_j + q_j \sigma^\mu \partial_\mu {q}^\dag_i \right)
\right] + h.c. .
\end{align}
Here the fermions are written in two-component Weyl fermions $q_i$, $\tilde{S}$. 
Kinetic terms for the SM fermions are induced in the last line.  
With the charge assignment in Eq.~\eqref{eq-npat}, 
the kinetic terms, e.g. $C_Q^{ij} q_i^\dag \ol{\sigma}^\mu \partial_\mu q_j$, have the following texture, 
\begin{align}
 C_Q \sim 
\begin{pmatrix}
 1 & \eps^2 & \eps \\
\eps^2 & 1 & \eps  \\ 
\eps & \eps & 1 
\end{pmatrix},
\quad
C_d \sim 
\begin{pmatrix}
 1 & \eps & \eps \\
\eps & 1 & \eps^2  \\ 
\eps & \eps^2 & 1 
\end{pmatrix},
\quad 
 C_u \sim C_e \sim C_L \sim 
\begin{pmatrix}
 1 & \eps^2 & \eps^2 \\
\eps^2 & 1 & \eps^2 \\ 
\eps^2 & \eps^2 & 1 
\end{pmatrix}, 
\end{align}
The canonically normalized basis is obtained by redefining the fermions as $f \to f^\prime_i := P_f^{ij} f_j$, 
where $f = Q,u,d,L,e$. 
Here, $P_f$'s have the same texture as $C_f$'s.   
We find that 
this rescaling keeps the texture of the Yukawa matrices in Eq.~\eqref{eq-nYsamp},
changing only ${\cal O}(1)$ factors per order.

The first line in Eq.~\eqref{eq-highq} directly gives various four fermi operators 
which can induce flavor violation. 
Let us study two observables, namely $\br{\mu}{e \ol{e} e}$ and $\eps_K$, 
which may give the strongest limits for lepton and quark sectors, respectively.  
For simplicity, in the discussions below, we ignore off-diagonal elements of the unitary matrices for diagonalizing the mass matrices. 
The LFV decay, $\mu \to e \ol{e} e$ is induced by a operator,  
\begin{align}
\label{eq-LLLL}
  \int d^4 \theta &\ \frac{e_{L_2 L_1 L_1 L_1}}{\Lambda^2}  L_{L_2}^\dag L_{L_1} L_{L_1}^\dag L_{L_1}  
\supset 
- \frac{e_{L_2L_1L_1L_1}}{2\Lambda^2} \   
       \ol{\mu} \gamma^\mu P_L e \cdot \ol{e} \gamma_\mu P_L  e.    
\end{align}
Here, the fermions in the right-hand side are Dirac fermions. 
Note that the other operators are more suppressed by $\Lambda$ due to the $\Zn{4}^F$ symmetry.  
 A branching fraction coming from this operator is given by~\cite{Kuno:1999jp,Okada:1999zk}  
\begin{align}
 \br{\mu}{e \ol{e} e} \sim &\ 
       \frac{m_{\mu}^5}{3072\pi^3 \Gamma_{\mu} } \frac{\abs{e_{L_2L_1L_1L_1}}^2}{\Lambda^4}  \\ \notag  
 \sim&\  7\times 10^{-15} \times \left(\frac{\abs{e_{L_2L_1L_1L_1}}}{1.0} \right)^2 
   \left(\frac{\eps}{0.02}\right)^4 \left(\frac{10\ \mathrm{TeV}}{v_s} \right)^4.    
\end{align}   
This can be consistent with the current experimental bound.

As another example, a four fermi operator relevant to the $K$-$\ol{K}$ mixing is given by  
\begin{align}
\label{eq-dddd}
  \int d^4 \theta &\ \frac{e_{Q_{2} Q_{1} \ol{d}_{1} \ol{d}_{2}}}{\Lambda^3} 
             S Q_{L_2}^\dag Q_{L_1} \ol{d}_{R_1}^\dag \ol{d}_{R_2}  
\supset 
 \frac{\eps  e_{Q_2Q_1 \ol{d}_1\ol{d}_2  }}{2 \Lambda^2} \   
                 \ol{s} \gamma^\mu P_L d \cdot \ol{s} \gamma_\mu P_R  d.     
\end{align} 
This will give the largest contribution due to the larger hadronic matrix elements 
of the left-right type operators~\cite{Aoki:2019cca}.    
The size of a contribution to $\eps_K$ is estimated as 
\begin{align}
\abs{\Delta \eps_K} =&\  \frac{\kappa_\eps}{\sqrt{2} \Delta M_K} 
                            \frac{\eps \cdot \mathrm{Im} \left( e_{Q_2Q_1\ol{d}_1\ol{d}_2}\right) }{2\Lambda^2} 
                        \abs{ \mathcal{O}_1^{\mathrm{LR}}  }  \\ \notag 
\sim&\ 10^{-2} \times \left(\frac{\eps}{0.02}\right)^3 \left(\frac{100\ \mathrm{TeV}}{v_s} \right)^2 
                        \left( \frac{\mathrm{Im} \left(e_{Q_2Q_1\ol{d}_1\ol{d}_2}\right) }{1.0}  \right).  
\end{align} 
The value of the hadronic matrix element $\mathcal{O}_1^{LR}$ is 
shown in Table~\ref{tab-valOs}. 
This is bigger than the experimental value
$\eps_K = 2.228\times10^{-3}$~\cite{Tanabashi:2018oca}
by one order of magnitude. 
For this correction to be consistent with the experimental value, 
$\mathrm{Im} \left( e_{Q_2Q_1\ol{d}_1\ol{d}_2} \right) \sim \order{0.01}$ 
is required unless $v_s\gtrsim 1$ PeV.
Hence, for the Higgsino LSP case,
$e_{Q_2Q_1\ol{d}_1\ol{d}_2}$ itself should be suppressed 
or aligned with the phase of the SM contribution for some reasons.  
For the wino LSP case with $v_s\gtrsim 1$ PeV, this problem can be evaded.
At any rate, the origin of this operator depends on UV physics.
In addition to the four fermi operators in the first line,
the second line of Eq.~\eqref{eq-highq} also induces four fermi operators by the flavon exchanging. 
However, these are more suppressed by a ratio of fermion to flavon mass than those from the first line. 
Hence, the four fermi operator from the first line would be the dominant one. 
This type of \Kahler\ potential will have various combinations of four fermi operators, 
and then affect to various flavor violating observables.

In summary,
the higher dimensional operators in the \Kahler\ potential 
will not change the hierarchical structure of the Yukawa matrices and the 
Higgs potential, but will only affect to $\order{1}$ factors per order.
On the other hand, 
these can induce new flavor violating effects  
and would put strong lower bounds on the flavon VEV $v_s$. 
However, it depends on how the operators are realized in an UV completion of this model.   
In the main text of this paper, we studied contributions which always exist 
as long as the Yukawa hierarchy is explained by the superpotential Eq.~\eqref{eq-Wyuk}. 
Note that the Yukawa hierarchies and the Higgs potential are not changed 
even if the cutoff scale is so large that the flavor violating effects are 
sufficiently suppressed. 
Potential problems of a large cutoff scale will be a relic density of the LSP 
and the 125 GeV Higgs boson mass. 
Detailed study of the higher dimensional operators in the \Kahler\ potential 
is left as our future work and will be discussed together with the UV completion of this model.

\section{Numerical coefficients}
\label{App-Bench}  
In this paper, we assume $\eps = 0.0195764 = \left(m_u/m_t\right)^{1/3}$.
The singular values of Yukawa matrices $Y^f$ 
(square roots of eigenvalues of $Y^{f\dag} Y^f$ or $Y^f Y^{f\dag}$) are fitted 
to the values at 1 TeV~\cite{Antusch:2013jca}. 
For the Yukawa couplings and Majorana neutrino masses, 
we used $\order{1}$ coefficients (of absolute value)  lying in the range of  $ [0.579, 7.11]$ as below:
\begin{align}
c^u 
=&\ 
\left(
\begin{array}{ccc}
-2.23656 & -3.78792 & 5.07947\cdot e^{-2.23037i} \\
 -1.8029 & 1.51612 & -0.62796 \\
 2.43468\cdot e^{0.019714 i} & -2.11793 & 0.782311 \\
\end{array}
\right), \\
c^d
=&\ 
\left(
\begin{array}{ccc}
 7.11034 & 4.75778 & 4.38956 \cdot e^{-1.64741i } \\
 6.74255 & -5.32201 & 3.39087 \\
 2.85434\cdot e^{2.96002i} & -0.578767 & -2.59023 \\
\end{array}
\right), \\
c^e
=&\  
\left(
\begin{array}{ccc}
 -1.83414 & -4.06715 & -4.55088 \\
 0.814655 & -1.04839 & -1.16518 \\
 -0.702312 & 1.27439 & 1.27222 \\
\end{array}
\right). \\ 
c^n
=&\ 
\left(
\begin{array}{ccc}
 3.63525 & -4.36595 & -4.00992 \\
 -5.94856 & -2.38206 & 3.74011 \\
 -2.19846 & -1.4343 & 0.589928 \\
\end{array}
\right), \\
M
 =&\   M_0 
\left(
\begin{array}{ccc}
 -6.07582 & 2.75669 & 4.32291 \\
 2.75669 & -4.43903 & 1.68412 \\
 4.32291 & 1.68412 & 5.09895 \\
\end{array}
\right),  
\end{align}
where $M_0$ is an overall scale of the Majorana mass. 
These values together with the hierarchical structure Eq.~\eqref{eq-nYsamp}  
lead to the fermion masses (in unit of [GeV]) and CP phases of CKM matrix 
\begin{align}
 (m_u, m_c, m_t) =&\ (0.001288, 0.6268, 171.7),
\quad 
 (m_d, m_s, m_b)= (0.002751, 0.05432, 2.853), \notag \\
 (m_e,m_\mu, m_\tau) =&\ (0.0004866, 0.1027, 1.746),  
\quad
(\alpha_{\rm CKM},\sin2\beta_{\rm CKM}, \gamma_{\rm CKM}) 
=  (1.518,  0.6950, 1.240), \notag     
\end{align}
and the absolute values of CKM matrix  
\begin{align}
 \abs{V_\mathrm{CKM}} = 
\left(
\begin{array}{ccc}
 0.974461 & 0.224529 & 0.00364284 \\
 0.224379 & 0.97359 & 0.0421456 \\
 0.00896391 & 0.0413421 & 0.999105 \\
\end{array}
\right),
\end{align}
where $\tan\beta = 5$. 
With $M_0 = 33.1474$ TeV and $\ell=3$, 
the neutrino mass differences (in unit of [eV$^2$]) are 
\begin{align}
\Delta m_{12}^2 =  7.37\times 10^{-5},\quad \Delta m_{23}^2 = 2.56\times 10^{-3}, 
\end{align} 
and the PMNS angles are 
\begin{align}
 \sin^2 \theta_{12} = 0.297,\quad 
 \sin^2 \theta_{23} = 0.425,\quad 
 \sin^2 \theta_{13} = 0.0215. 
\end{align}
In this model, the Majorana mass may naturally be given by the cutoff scale,
\begin{align}
 \Lambda \sim 500\ \mathrm{TeV} \times \left(\frac{0.02}{\eps} \right) \left(\frac{v_s}{10\ \mathrm{TeV}} \right).  
\end{align}

{\small 
\bibliographystyle{JHEP}
\bibliography{reference_EWFN}
}
\end{document}